\documentclass[reprint,nofootinbib,amsmath,amssymb,aps]{revtex4-2}

\usepackage[table]{xcolor}
\usepackage{mathtools}
\usepackage{bm}
\usepackage{dsfont}
\usepackage{braket}

\begin{document}

\title{Generation of Scalable Genuine Multipartite Gaussian Entanglement\\ with a Parametric Amplifier Network}
\author{Saesun Kim$^{1,2}$}
\email{saesun.kim@jpl.nasa.gov}
\author{Sho Onoe$^{3,4}$}
\author{Alberto M. Marino$^{1,2,5,6}$}
\email{marino@ou.edu, marinoa@ornl.gov}
\affiliation{$^{1}$Homer L. Dodge Department of Physics and Astronomy, The University of Oklahoma, Norman, OK 73019, USA}
\affiliation{$^{2}$Center for Quantum Research and Technology, University of Oklahoma, Norman, OK, 73019, USA}
\affiliation{$^{3}$Centre for Quantum Computation and Communication Technology, School of Mathematics and Physics,
The University of Queensland, St. Lucia, Queensland, 4072, Australia}
\affiliation{$^{4}$femtoQ Lab, Department of Engineering Physics, Polytechnique Montr\'{e}al, Montr\'{e}al, QC H3T 1JK, Canada}
\affiliation{$^{5}$Quantum Information Science Section, Computational Sciences and Engineering Division, Oak Ridge National Laboratory, Oak Ridge, TN 37831, USA}\thanks{This manuscript has been authored in part by UT-Battelle, LLC, under contract DE-AC05-00OR22725 with the US Department of Energy (DOE). The
publisher acknowledges the US government license to provide public access under the DOE Public Access Plan (http://energy.gov/downloads/doe-publicaccess-plan).}
\affiliation{$^{6}$Quantum Science Center, Oak Ridge National Laboratory, Oak Ridge, TN 37381, USA}
\begin{abstract}
Genuine multipartite entanglement is a valuable resource in quantum information science, as it exhibits stronger non-locality compared to bipartite entanglement. This non-locality can be exploited in various quantum information protocols, such as teleportation, dense coding, and quantum interferometry. Here, we propose a scheme to generate scalable genuine multipartite continuous-variable entangled states of light using a parametric amplifier network. We verify the presence of genuine quadripartite, hexapartite, and octapartite entanglement through a violation of the positive partial transpose (PPT) criteria. Additionally, we use $\alpha$-entanglement of formation to demonstrate the scalability of our approach to an arbitrary number of $2N$ genuinely entangled parties by taking advantage of the symmetries present in our scheme.
\end{abstract}

\maketitle

\section{\label{Intro}Introduction}

Genuine multipartite entanglement (GME) is the strongest form of non-locality that a quantum state with a given number of parties can have. It is characterized by entanglement among all possible bi-partitions of the modes that make up the state. Given this high degree of non-locality, it is considered  a valuable resource for quantum information science~\cite{PhysRevApplied.13.054022} and has been proposed as a resource for teleportation and dense coding~\cite{densec}, to enhance the rate of quantum key distribution~\cite{QKD}, and to obtain the highest possible sensitivities for quantum interferometry~\cite{PhysRevA.85.022322,PhysRevA.85.022321}.

Most of the work to date on the generation of multipartite entanglement has focused on discrete variable (DV) systems, in which information can be encoded in the discrete eigenstates of the system, such as the two-dimensional Hilbert space of the polarization of a photon or the spin of a particle~\cite{PhysRevLett.121.250505,Pueaar3931,PhysRevLett.119.180511,mooney_hill_hollenberg_2019}. Another possibility is the use of continuous variable (CV) systems, which are characterized by a continuous distribution of eigenstates that can be used to encode information, such as the amplitude and phase quadratures of the field. CV systems offer significant advantages, as they allow for the efficient realization of quantum information protocols~\cite{ef1,ef2,ef3}, offer an efficient interaction with atomic ensembles~\cite{atomSq1,atomSq2}, and can deterministically generate large-scale multipartite entanglement~\cite{oliver,million}. GME in CV systems has been realised experimentally and approaches to scale up to larger number of parties through the use of an array of beam splitters~\cite{Reid,Furusawa} or arrays of parametric amplifiers~\cite{cascade,cascade2,Misra_2022} have been proposed. However, as the number of parties increases, the structure of the correlations between the different modes becomes more complex.

Here, we propose a new scheme for generating genuine multipartite entangled states of light  based on a parametric amplifier network. The proposed scheme consists of two stages of parametric amplifiers and does not require an array of optical elements, which significantly reduces the complexity of the system. The resulting entanglement structure is simple and symmetric, making it easy to characterize the correlations between  all the entangled modes. This makes the proof of GME when extending to a large number of parties tractable.

We first show that the proposed parametric amplifier network can generate genuine quadripartite, hexapartite, and octapartite entanglement through a violation of the positive partial transpose (PPT) criteria~\cite{Lami_2018}. By verifying the inseparability of all possible bi-partitions of the system, we gain insight into the structure of the correlations and the symmetries of the generated multipartite entangled state. We then utilize this understanding to simplify the system through a reduction method that involves unitary transformations, which makes it possible  to analytically prove the presence of GME. We do so through the use of $\alpha$-entanglement of formation to show that the Von Neumann entropy of all possible partitions is larger than or equal to the one of the reduced system. This shows that our system is scalable and is capable of generating genuine $2N$-partite entanglement.

\section{Parametric Amplifier Network}

\begin{figure}[h]
    \centering
    \includegraphics{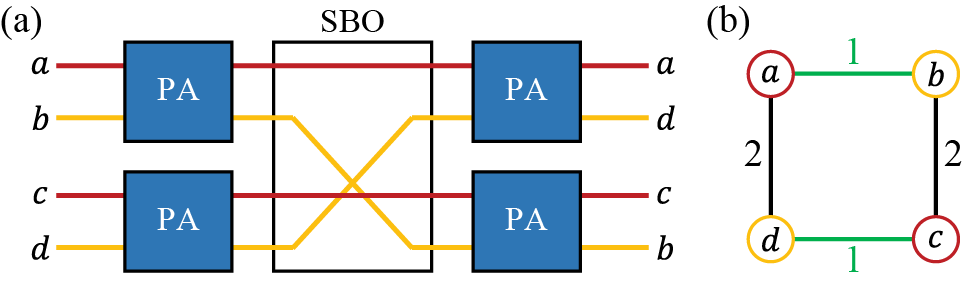}
    \caption{(a) Schematic of proposed PA network for the generation of genuine quadripartite entangled states. (b) Graphical representation of the generated quadripartite entangled state, with the green lines (labeled as 1) and the black lines (labeled as 2) representing the correlations introduced by the first or second stage of parametric amplifiers, respectively. PA: parametric amplifier; SBO: switchboard operation.}
\label{square}
\end{figure}

The proposed scheme is based on a two-stage network of parametric amplifiers connected through an optical system that routes the modes between the two stages in what we refer to as a switchboard operation. We start with a description of the system to extend from two entangled modes with a single parametric amplifier to four entangled modes with the proposed scheme. As shown in Fig.~\ref{square}(a), four modes \{$a$, $b$, $c$, $d$\} serve as inputs to the first stage of the network composed of two parametric amplifiers.  This first stage entangles mode pairs $a \leftrightarrow b$ and $c \leftrightarrow d$. After the first stage, the switchboard operation swaps modes $b$ and $d$ while directly transmitting modes $a$ and $c$. The outputs of the switchboard serve as the inputs for the second stage of the network, which is composed of two additional parametric amplifiers. The second stage entangles mode pairs  $a \leftrightarrow d$ and $c \leftrightarrow  b$. Figure~\ref{square}(b) shows the graphical representation of the connections between the four output modes with a square geometry. The green lines (labeled 1) and the black lines (labeled 2) correspond to the connections introduced by the first or second stage of parametric amplifiers, respectively.

To verify the presence of GME in this quadripatite system, we first construct the covariance matrix (CM). Given that the system is Gaussian, the CM provides a complete characterization of the quantum properties of the system.  The CM is determined by second-order moments of the form $\sigma=\left\langle \bm{\xi\xi}^T \right\rangle$, where $\bm{\xi}$ is the quadrature element vector $\bm{\xi}=[\hat{X}_a, \hat{Y}_a, \hat{X}_b, \hat{Y}_b, \hat{X}_c, \hat{Y}_c, \hat{X}_d, \hat{Y}_d]$ with the quadrature operators defined as $\hat{X}_a=(\hat{a}^\dagger+\hat{a}) $ and $\hat{Y}_a=i(\hat{a}^\dagger-\hat{a})$ in terms of the annihilation, $\hat{a}$, and creation, $\hat{a}^\dagger$, operators. Thus, the elements of the CM take the form $\left\langle \sigma_{ij}\right\rangle=\left\langle \hat{X}_i \hat{X}_j+ \hat{X}_j \hat{X}_i\right\rangle/2-\left\langle \hat{X}_i\right\rangle\left\langle \hat{X}_j\right\rangle$ for all possible combinations of quadratures and modes.

Since the field operators follow the bosonic commutation relation  $[\hat{a},\hat{a}^\dagger]=1$, the quadratures satisfy the commutator relation $[\hat{X}_a,\hat{Y}_a]=2i$. As a result, due to the uncertainty principle, a physical CM must satisfy the condition $\sigma+i \Omega\geq0$~\cite{Simon3}, where the symplectic matrix $\Omega$ is given by
\begin{equation}
    \Omega=\overset{N}{\underset{j=0}{\oplus}}J\qquad {\rm with}\qquad
    J=\left(\begin{matrix}
        0 & -1\\
        1 & 0
    \end{matrix}\right).
\end{equation}
Similarly, for a partially transposed CM, which we denote as $\widetilde{\sigma}$, this condition takes the form $\widetilde{\sigma}+i \Omega\geq0$, which is equivalent to $\Delta\equiv-(\widetilde{\Omega} \sigma)^2\geq1$ for a pure quantum state. This condition establishes a necessary and sufficient criterion for the separability of the system~\cite{Lami_2018}. Violation of this criterion determines that the system is inseparable, with a stronger violation indicating a stronger degree of entanglement. Thus, we can verify the presence of entanglement by calculating the smallest symplectic eigenvalue of matrix $\Delta$. We can confirm the presence of GME using the PPT criteria if it can be shown that all possible bi-partitions are inseparable~\cite{Reid}.

To construct the CM of the generated state, shown in Fig.~\ref{square}(b), we describe the operation performed by the parametric amplifiers with the two-mode squeezing operator $\hat{S}_{ab} (\zeta)=\textrm{exp}(\zeta^* \hat{a}\hat{b}-\zeta \hat{a}^\dagger \hat{b}^\dagger)$, where $\zeta=s~\textrm{exp}(i \theta)$ with $s$ and $\theta$ representing the degree of squeezing and the squeezing angle, respectively. We choose the squeezing angle that maximizes the entanglement in the system, which corresponds to $\theta=\pi$. The corresponding CM, given in App.~\ref{app:A}, is then used to obtain an analytical expression for the smallest symplectic eigenvalue for all possible bi-partitions.

\begin{figure}[h!]
  \centering
  \includegraphics{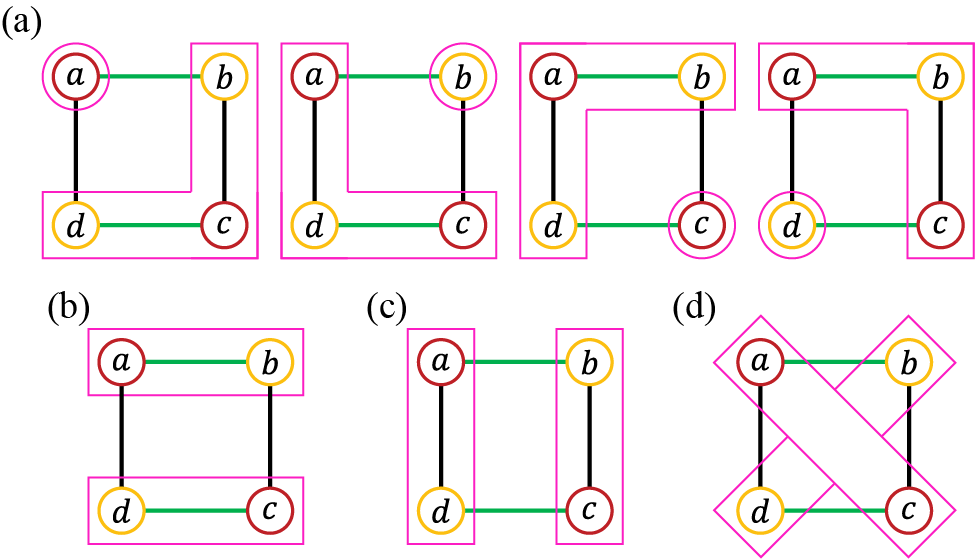}
  \caption{(a)~Graphical representation of all possible bi-partitions of the form 1$\times$3 and (b) through (d)  bi-partitions of the form of 2$\times$2.}
  \label{PPTgraph}
\end{figure}

\begin{figure}[t!]
  \centering
  \includegraphics{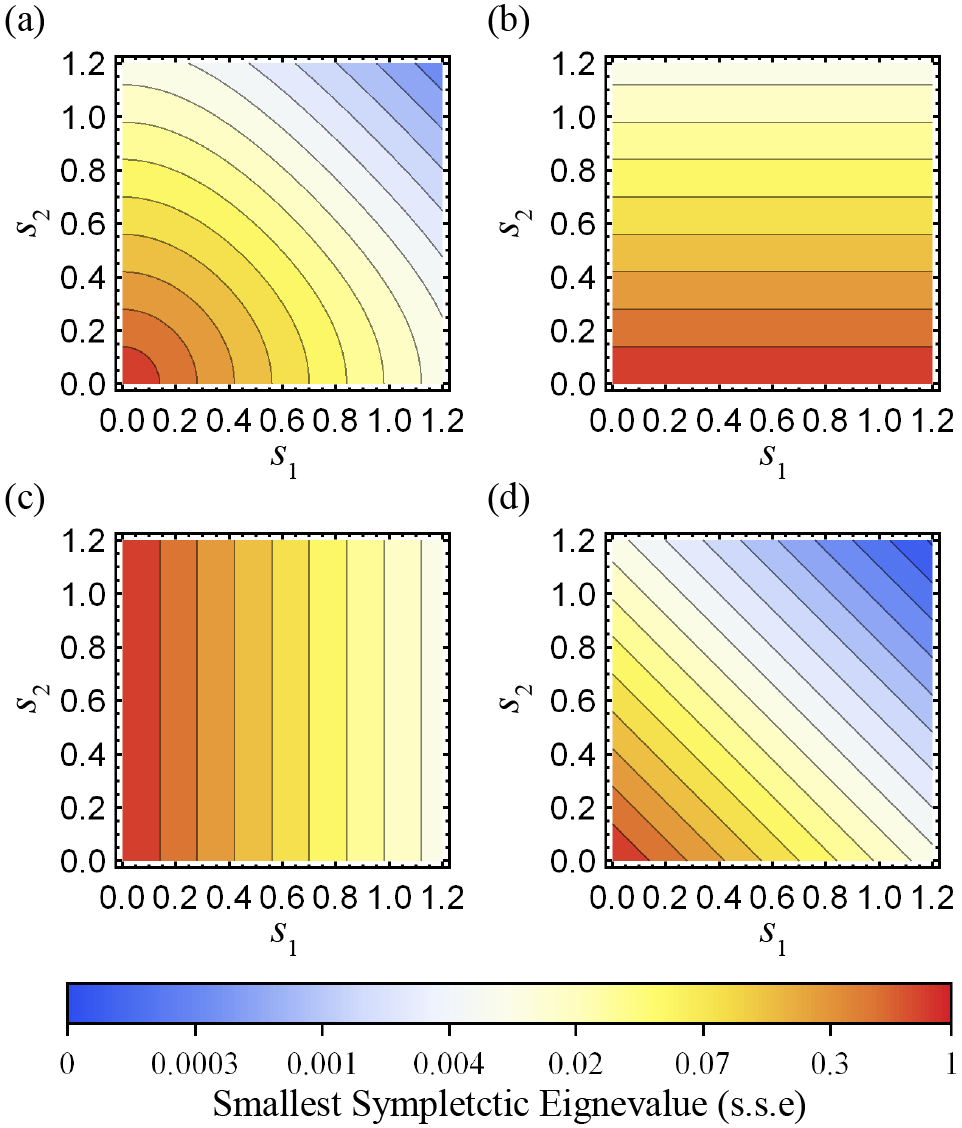}
  \caption{Contour plots of the smallest symplectic eigenvalues of the different bi-partitions with respect to the squeezing parameters of the two stages of the PA network. Values less than 1 show the inseparability for bi-partitions of the form (a) 1$\times$3 and of the form 2$\times$2 for grouping (b) $ab|cd$, (c) $ad|bc$, and (d) $ac|bd$.}
  \label{PPT}
\end{figure}

We start with bi-partitions of the form 1$\times$3, shown in Fig.~\ref{PPTgraph}(a), for which there are four possible groupings given by $a|bcd$, $b|acd$, $c|abd$, and $d|abc$. In this case the PPT criteria for all bi-partitions are identical due to the symmetry of the state and take the form
\begin{equation}
    \begin{split}
        \Delta^{(1\times3)}&=\alpha _-^2 \left(\beta _+^2-\beta _-^2\right)+\alpha _+^2 \left(\beta _-^2+\beta _+^2\right) \\
        &-2 \alpha _+ \beta _+ \sqrt{\alpha _-^2 \left(\beta _+^2-\beta _-^2\right)+\alpha _+^2 \left(\beta _-^2+\beta _+^2\right)-\alpha _+^2 \beta _+^2},
    \end{split}
\end{equation}
where $\Delta^{(1\times3)}$ is the smallest symplectic eigenvalue of the PPT matrix $\Delta$, $\alpha _{\pm} = (e^{-2 s_1}\pm e^{2 s_1})/2$, and $\beta _{\pm}= (e^{-2 s_2}\pm e^{2 s_2})/2$.  As the graphical representation in Fig.~\ref{PPTgraph}(a) shows, the 1$\times$3 bi-partitions have the same connections between the two partitions, as indicated by the green and black lines that corresponds to the first and second stage of PAs, respectively.  Figure~\ref{PPT}(a) shows a contour plot of the smallest symplectic eigenvalue, $\Delta^{(1\times3)}$, as a function of the squeezing parameters of the first and the second stages. As can be seen, $\Delta^{(1\times3)}<1$ as long as $s_1>0$ or $s_2>0$.

Next, we consider the bi-partitions of the form 2$\times$2, shown in Figs.~\ref{PPTgraph}(b) through~\ref{PPTgraph}(d), which correspond to groupings $ab|cd$, $ad|bc$, and $ac|bd$. For these bi-partitions, the smallest symplectic eigenvalues are all different and take the form $\left(\alpha _+^2-\alpha _-^2\right) \left(\beta _-+\beta _+\right)^2$ for $ab|cd$, $\left(\alpha _-+\alpha _+\right){}^2 \left(\beta _-+\beta _+\right)^2$ for $ad|bc$, and $\left(\alpha _-+\alpha _+\right)^2\left(\beta _+^2-\beta _-^2\right)$ for $ac|bd$.
The expressions for groupings $ab|cd$ and $ad|bc$ simplify to $e^{-2 s_2}$ and $e^{-2 s_1}$, respectively. Thus, they are independent of the first or second stage, as can be seen in Figs.~\ref{PPT}(b) and~\ref{PPT}(c), given that the two partitions are only connected by correlations introduced by either the second or first PA stage. On the other hand,  grouping $ac|bd$ depends on the gain of both stages, as shown in Fig.~\ref{PPT}(d), given that modes $a$ and $c$ are indirectly connected through both PA stages.  This makes it the most entangled bi-partition of the four-mode case.  As can be seen from Figs.~\ref{PPT}(b) through~\ref{PPT}(d), all the bi-partitions of the form 2$\times$2 have smallest symplectic eigenvalues less than one when $s_1>0$ and $s_2>0$.

For all bi-partition, in the limit in which $s_1\rightarrow 0$ and $s_2\rightarrow0$ (no parametric amplification in either stage of the network) we have that $\alpha _{+}=\beta _{+}\rightarrow 1$ and $\alpha _{-}=\beta _{-} \rightarrow 0$, such that the smallest symplectic eigenvalues all tend to 1. This implies that the generated state is separable when there is no squeezing from any of the PAs, as expected given that there are no interactions (through the PAs) between any of the four independent input modes. On the other hand, as soon as the squeezing parameters from both stages are greater than 0, the smallest symplectic eigenvalues of all possible bi-partition become less than one. These results  show that all the bi-partitions are inseparable, which implies that the generated state is a genuine quadrapartite entangled state.

\begin{figure}[b]
  \centering
  \includegraphics{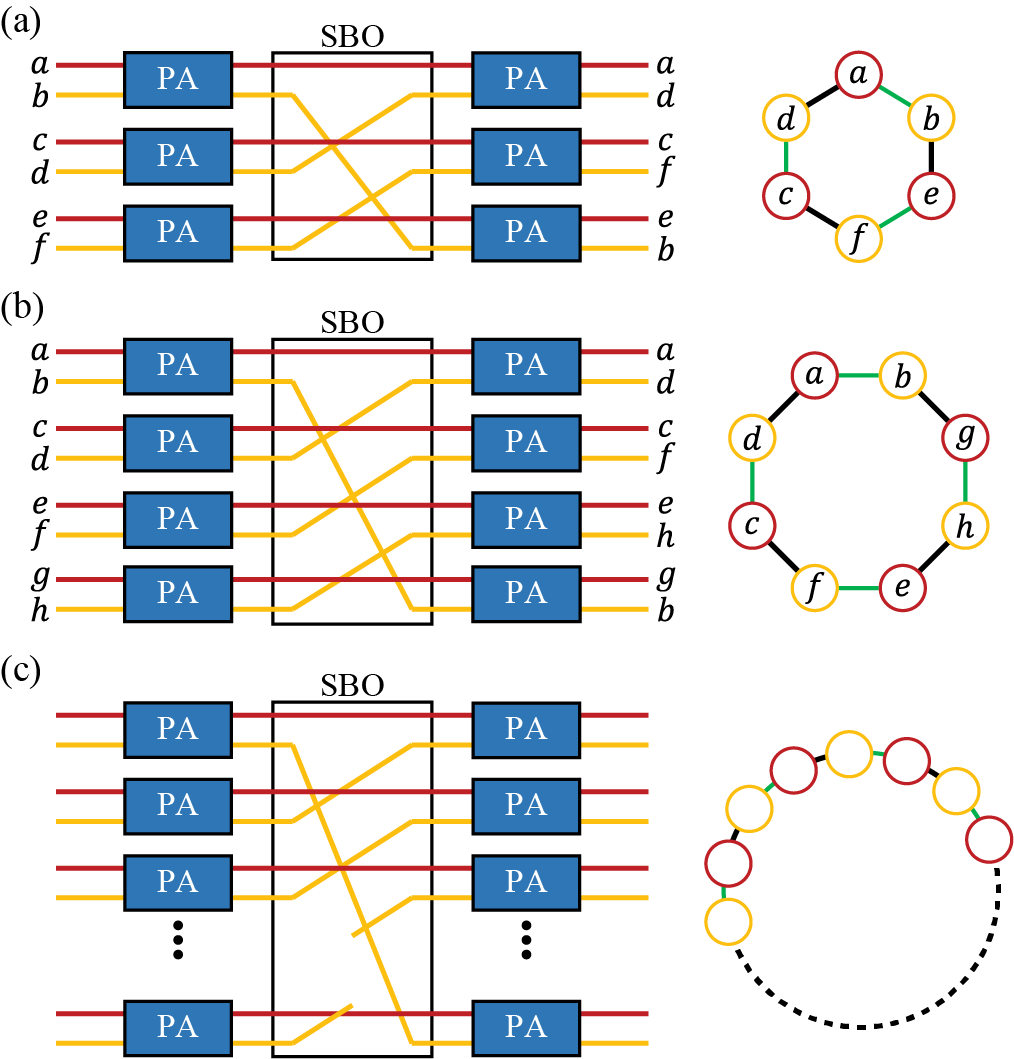}
  \caption{Schematic of the PA network for the generation of genuine (a) hexapartite and (b) octapartite entanglement. (c) Generalization of the PA network for the generation of genuine $2N$-partite entanglement. For all cases, the corresponding graphical representation of the generated state is shown to the right of the PA network used to generate it.  PA: parametric amplifier; SBO: switchboard operation.}
  \label{multibeam}
\end{figure}

\section{\label{HexOct}Genuine Hexapartite and Octapartite Entanglement}

\begin{figure*}
  \centering
  \includegraphics{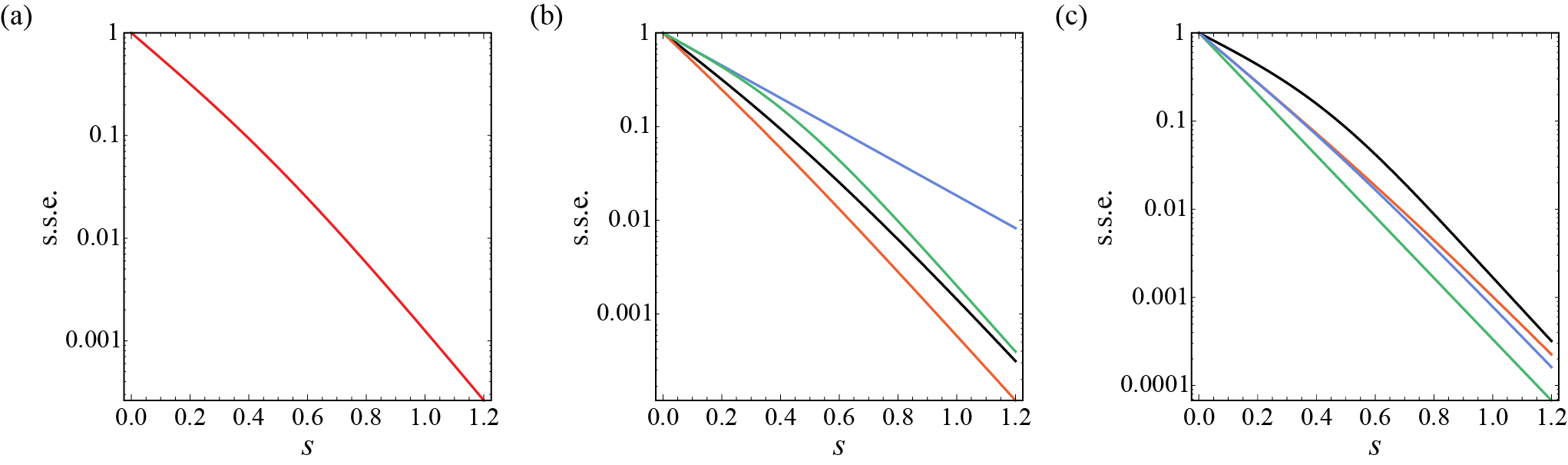}
  \caption{Smallest symplectic eigenvalue (s.s.e.) for bi-partitions of the form (a)~1$\times$5, (b)~2$\times$4, and (c)~3$\times$3 for the generated hexapartite entanglement state with respect to the squeezing parameters of the PA stages, assumed to be the same ($s_1=s_2=s$). A detailed description of the different traces for each case is given in the main text. All possible bi-partitions show a s.s.e less than 1 for a squeezing parameter greater than 0, which implies GME for the generated hexapartite state.}
  \label{PPT6}
\end{figure*}

We now show that through a careful choice of the switchboard operation and the addition of parametric amplifiers in each stage, it is possible to extend the proposed scheme to generate genuine hexapartite and octapartite entanglement. Figures~\ref{multibeam}(a) and~\ref{multibeam}(b) show the proposed schemes for their generation and the corresponding graphical representation of the states.

As the number of modes increases, the total number of possible bi-partitions grows rapidly. For an $N$-mode system, the total number of bi-partitions that need to be considered to exhibit a violation of the PPT criteria is given by $2^{N-1}-1$~\cite{Devlin_1979}. Furthermore, for bi-partitions of the form $k \times (N-k)$, the possible number of groupings can be calculated through the binomial coefficient
\begin{equation}
    \prescript{N\mkern-0.5mu}{}C_{k}=\frac{N(N-1)(N-2)\cdots (N-k+1)}{k(k-1) \cdots 1}.
\end{equation}
For example, for the hexapartite system there are a total of $2^5-1=31$ bi-partitions that need to be verified for a violation of the PPT criteria. More specifically, for bi-partitions of the form 1$\times$5 there are $\prescript{6\mkern-0.5mu}{}C_{1}=6$ groupings, for bi-partitions of the form 2$\times$4 there are $\prescript{6\mkern-0.5mu}{}C_{2}=15$ groupings, and for bi-partition of the form 3$\times$3 there are $\prescript{6\mkern-0.5mu}{}C_{3}/2=10$ groupings. Note that for bi-partition of the form 3$\times$3 it is necessary to divide the binomial coefficient by two to prevent double counting.  In what follows, we use the SNEG package~\cite{ZITKO} in Mathematica to calculate the smallest symplectic eigenvalues for all possible bi-partions for a hexapartite and an octapartite system.

\begin{figure}[hb!]
  \centering
  \includegraphics{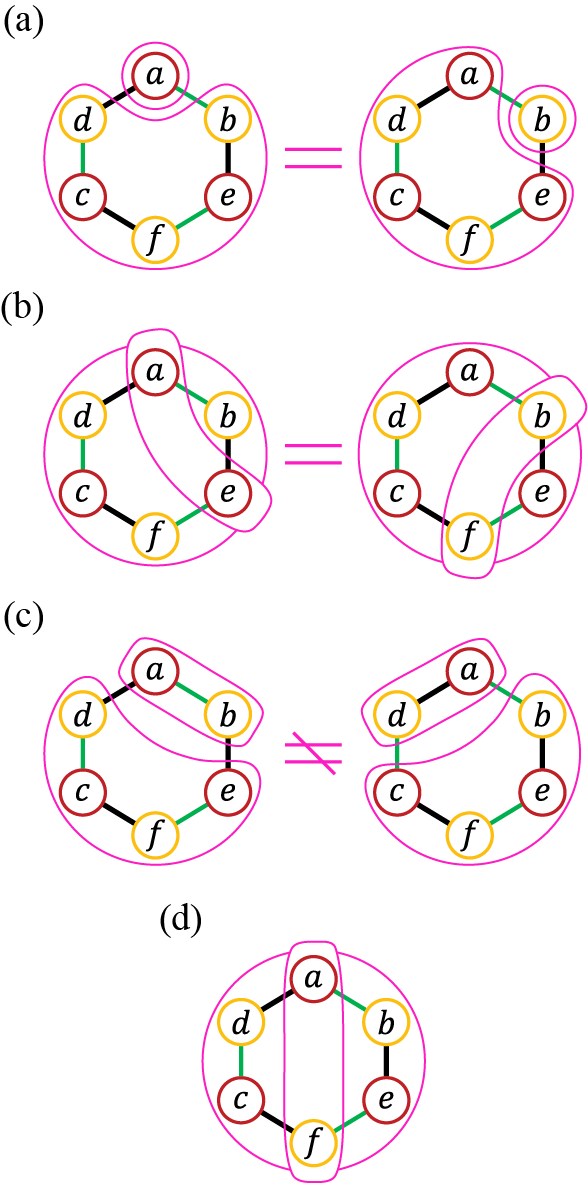}
  \caption{(a)~Graphical representation of bi-partition of form 1$\times$5. (b)~through~(d)~Graphical representations of bi-partition of form 2$\times$4.}
  \label{PPT6-1&2}
\end{figure}

We start by considering the hexapartite system shown in Fig.~\ref{multibeam}(a). For bi-partitions of the form 1$\times$5 the possible groupings are $a|bcdef$, $b|acdef$, $c|abdef$, $d|abcef$, $e|abdf$, and $f|abcde$. Figure~\ref{PPT6}(a) shows the smallest symplectic eigenvalues for these bi-partitions for the case in which both stages of the PA network have the same squeezing parameter ($s_1=s_2=s$).  As can be seen, all the smallest symplectic eigenvalues are identical. This is expected due to the symmetry of the connections between the resulting partitions, as shown schematically in Fig.~\ref{PPT6-1&2}(a). The smallest symplectic eigenvalues without the assumption of equal squeezing in both stages of the PA netowrk are shown in App.~\ref{app:B}.

\begin{figure}[hb!]
  \centering
  \includegraphics{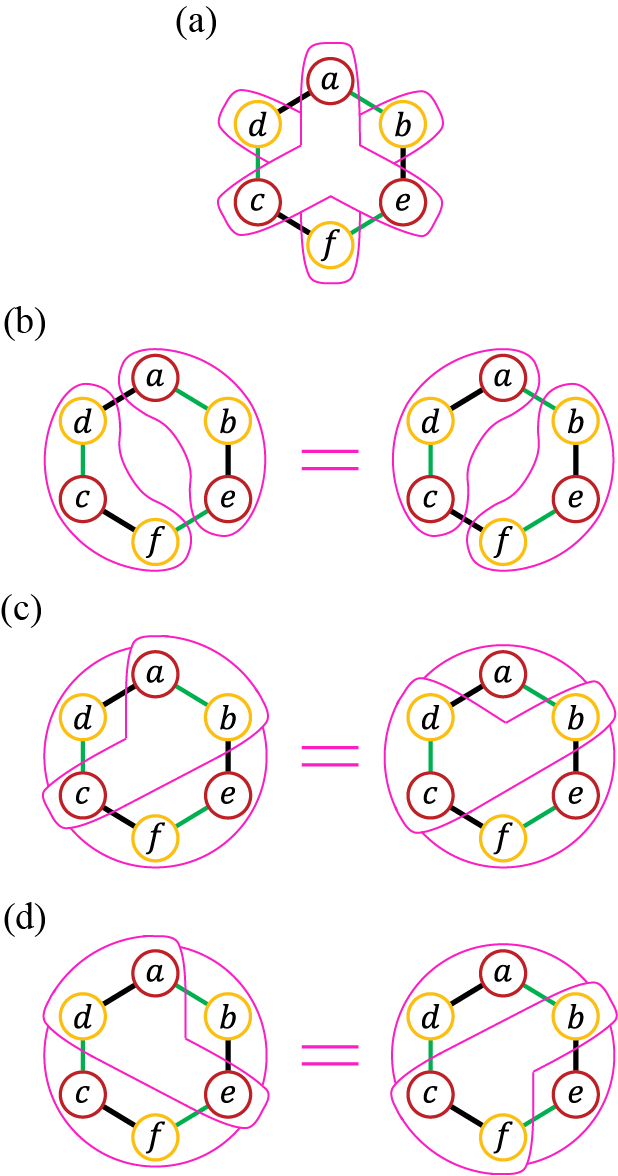}
  \caption{(a)~through~(d)~Graphical representations of bi-partitions of the form  3$\times$3.}
  \label{PPT6-3}
\end{figure}

Bi-partitions of the form 2$\times$4 and 3$\times$3 are more complex, as can be seen  by the smallest symplectic eigenvalues shown in Figs.~\ref{PPT6}(b) and~\ref{PPT6}(c), respectively. For bi-partitions of the form 2$\times$4 there are four unique cases that are shown schematically in Figs.~\ref{PPT6-1&2}(b) through~\ref{PPT6-1&2}(d). The bi-partitions between second nearest neighbors and the rest of the system ($ae|bcdf$, $bf|acde$, $ce|abdf$, $df|abce$, $ac|bdef$, $bd|acef$) show the strongest violation of the PPT criteria, as can be seen by the orange trace in Fig.~\ref{PPT6}(b).  This indicates that these bi-partitions, represented in Fig.~\ref{PPT6-1&2}(b), show the strongest entanglement for bi-partitions of the form 2$\times$4. For this bi-partitioning of the system, modes such as $a$ and $e$ are indirectly connected through one PA from the first stage and one from the second stage. Since these are the bi-partitions that group modes that are the closest possible while breaking connections from both PA stages, they have the strongest possible correlations. Indirect connections through more or less PA processes will make the correlations between the modes weaker.

The bi-partitions with the second-to-highest degree of entanglement, given by the black trace in Fig.~\ref{PPT6}(b), correspond to bi-partitions in which the elements of one of the partitions are indirectly connected by two PAs from the first stage and one from the second stage or by two from the second stage and one from the first stage ($af|bcde$, $bc|adef$, $de|abcf$), as shown in Fig.~\ref{PPT6-1&2}(d). While these bi-partitions break connections from both PA stages, the two modes are indirectly connected through more PAs that the bi-partitions shown in Fig.~\ref{PPT6-1&2}(b) and thus exhibit weaker entanglement.

The bi-partitions that shown the lowest degree of entanglement, given by the green and blue traces in Fig.~\ref{PPT6}(b), are those for which only connections from a single parametric amplifier stage are broken.  These correspond to bi-partitions between nearest neighbors and the rest of the system, with the neighbors connected by a first stage PA ($ab|cdef$, $ef|abcd$, $cd|abef$) or a second stage PA ($ad|bcef$, $be|acdf$, $cf|abde$), as shown schematically in Fig.~\ref{PPT6-1&2}(c). For these cases, partitions that have elements with nearest neighbors with a first stage connection, left diagram in Fig.~\ref{PPT6-1&2}(c), show more entanglement than ones with a second stage connection, right diagram.

This can be understood by considering the operations that connect the modes in  the partition with two elements.  For the left diagram, which corresponds to a bi-partitioning with one of the partitions composed of modes $a$ and $b$, the modes are connected through an entangling operation in the first stage followed by the addition of amplification noise in the second stage. In contrast, for the right diagram, which corresponds to a bi-partitioning with one of the partitions composed of modes $a$ and $d$, the modes result from an initial addition of amplification noise in the first stage followed by an entangling operation in the second stage. As demonstrated in App.~\ref{A:app4}, entangling first and amplifying later leads to a larger level of entanglement between the two partitions.

As can be seen in Fig.~\ref{PPT6}(b), the green and blue traces initially follow the same trend as the degree of squeezing of the PAs increases; however, the green trace then tends toward  the black one. This is due to the competition between the correlations introduced by both PA stages. As the degree of squeezing increases, the entanglement introduced by the second stage becomes stronger than the one introduced by the first stage. Therefore, in the large squeezing limit, bi-partitions that break connections introduced by the second stage ($ab|cdef$, $ef|abcd$, $cd|abef$) exhibit more entanglement and tend toward bipartitions of the form ($af|bcde$, $bc|adef$, $de|abcf$) given that the correlations introduce by the second stage dominate over the ones introduced by the first stage.

\begin{figure*}[t!]
  \centering
  \includegraphics{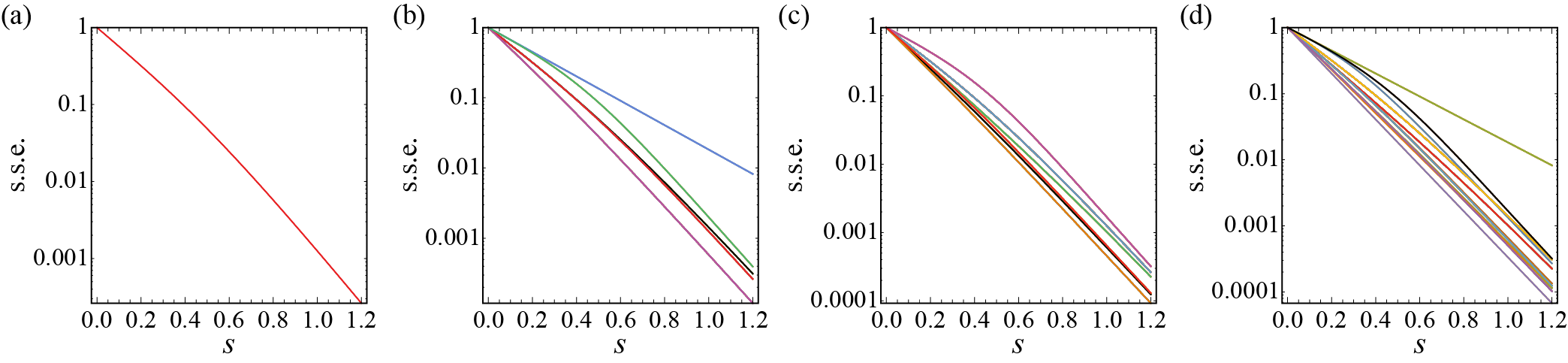}
   \caption{Smallest symplectic eigenvalues (s.s.e.) for the octapartite state with respect to squeezing parameter ($s_1=s_2$) for bi-partitions of the form (a)~1$\times$7, (b)~2$\times$6, (c)~3$\times$5, and (d)~4$\times$4. All possible bi-partitions show a s.s.e. less than 1 for a squeezing parameter greater than 0, which implies GME for the generated octapartite state.}
  \label{PPT8}
\end{figure*}

For bi-partitions of the form 3$\times$3, there are also four unique cases, as shown in Fig.~\ref{PPT6}(c) and represented schematically in Fig.~\ref{PPT6-3}. In this case, partitions consisting if only second nearest neighbor ($ace|bdf$, $bdf|ace$) have the largest degree of entanglement, as shown by the green trace in Fig.~\ref{PPT6}(c).  Following the same argument as with the bi-partitions of the form 2$\times$4, these bi-partitions group modes that are the closest possible while breaking the connections introduced by the first and the second stages, as shown in Fig.~\ref{PPT6-3}(a).

The bi-partitions with the second and third largest degrees of entanglement, given by the the blue and orange traces in Fig.~\ref{PPT6}(c), are ones that group two modes with nearest neighbors connected by a first stage PA with one mode that is a second nearest neighbor ($abc|def$, $abf|cde$, $aef|bcd$), as shown in Fig.~\ref{PPT6-3}(c), and ones that group two modes with nearest neighbors connected by a second stage PA with one mode that is a second nearest neighbor ($ade|bcf$, $adf|bce$, $bde|acf$), as shown in Fig.~\ref{PPT6-3}(d). There is again a difference between groupings that include two modes connected by a first stage PA or by a second stage PA, with ones with  nearest neighbors from the first stage having more entanglement than ones with nearest neighbors from the second stage. Finally, the bi-partitions with the least degree of entanglement, given by the black trace in Fig.~\ref{PPT6}(c), are ones with partitions composed of only nearest neighbors ($abe|cdf$, $bef|acd$, $cef|abd$). While these bi-partitions break connections from both parametric stages, as seen in Fig.~\ref{PPT6-3}(b), they break the least number of connections for all bi-partitions of the form 3$\times$3. As can be seen in Fig.~\ref{PPT6}, all possible bi-partitions needed to evaluate the PPT criteria have their smallest eigenvalue less than 1 when the squeezing is larger than 0, which implies GME for the generated hexapartite state.

Following the same trend, we can increase the number of entangled modes by adding an additional PA to the first and second stages, as shown in Fig.~\ref{multibeam}(b), to create genuine octapartite entanglement. For the octapartite system there are a total of $2^{8-1}-1=127$ possible bi-partitions that need to be verified. Figures~\ref{PPT8}(a),~\ref{PPT8}(b),~\ref{PPT8}(c), and~\ref{PPT8}(d) show the smallest symplectic eigenvalues for bi-partitions of the form 1$\times$7, 2$\times$6, 3$\times$5, and 4$\times$4, respectively. For bi-partitions of the form 1$\times$7, there are $\prescript{8\mkern-0.5mu}{}C_{1}=8$ cases that all exhibit the same behavior, as shown in Fig.~\ref{PPT8}(a), due to symmetry.

\begin{figure}[t!]
  \centering
  \includegraphics{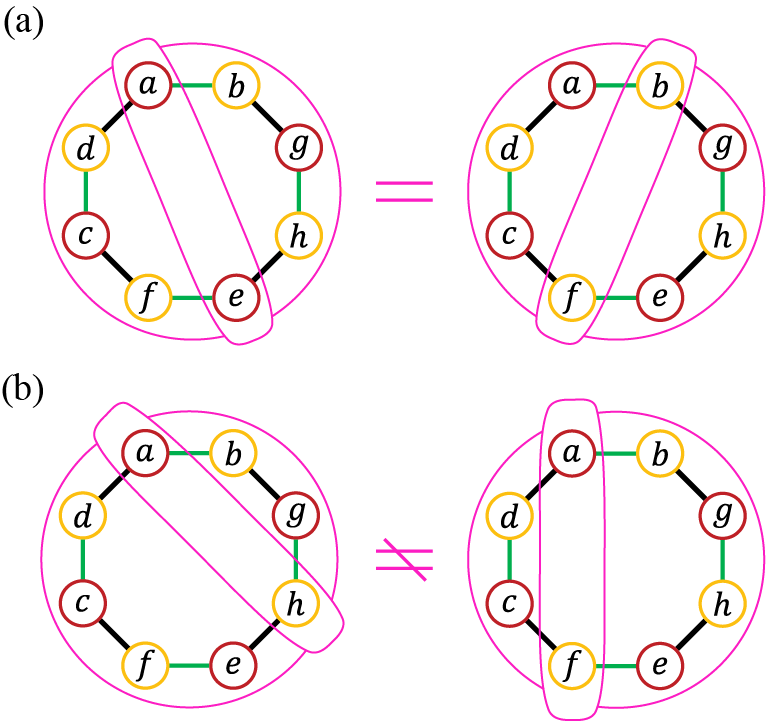}
  \caption{(a) and (b)~Graphical representations of bi-partitions of the form 2$\times$6.}
  \label{PPT8-1}
\end{figure}

Bi-partitions of the form 2$\times$6 for the octapartite system behave relatively similar to bi-partitions of the form 2$\times$4 for the hexapartite system. There are total of $\prescript{8\mkern-0.5mu}{}C_{2}=15$ partitions, but only 5 unique behaviors, as shown in Fig.~\ref{PPT8}(b). The smallest symplectic eigenvalues for different forms of bi-partitions for the octapartite case, shown in Fig.~\ref{PPT8}(b), and for the hexapartite case, shown in Fig.~\ref{PPT6}(b), follow the same ordering except for the one with the second-to-highest degree of entanglement, given by the red trace in Fig.~\ref{PPT8}(b). The red trace shows the behavior of the smallest symplectic eigenvalues for bi-partitions between fourth nearest neighbors and the rest of the system ($ae|bcdfgh$, $bf|acdegh$, $cg|abdefh$, $dh|abcefg$), as shown in Fig.~\ref{PPT8-1}(a), and ones between third nearest neighbors connected by two first stages and one second stage and the rest of the system ($ah|bcdefg$, $de|abcfgh$, $gf|abcdeh$, $bc|adefgh$), as shown in the left diagram in Fig.~\ref{PPT8-1}(b). The groupings between fourth nearest neighbors, shown in Fig.~\ref{PPT8-1}(a), are all equivalent due to the reflection symmetry of the state.

Since fourth nearest neighbors are always independent, third nearest neighbors and fourth nearest neighbors exhibit the same level of entanglement, which is a new feature of the octapartite system as the quadrapartite and the hexapartite systems do not have any two modes that come from independent processes. This leads to the same degree of entanglement between the left diagram in Fig.~\ref{PPT8-1}(b) and the ones in Fig.~\ref{PPT8-1}(a). That is, despite the fact that the distance between the two modes is different, they show the same inseparability. As we will discuss in the next section, this simplifies the structure of the correlations introduced by the system and makes it  easier to track them as the number of modes increase.

\begin{figure}[t!]
  \centering
  \includegraphics{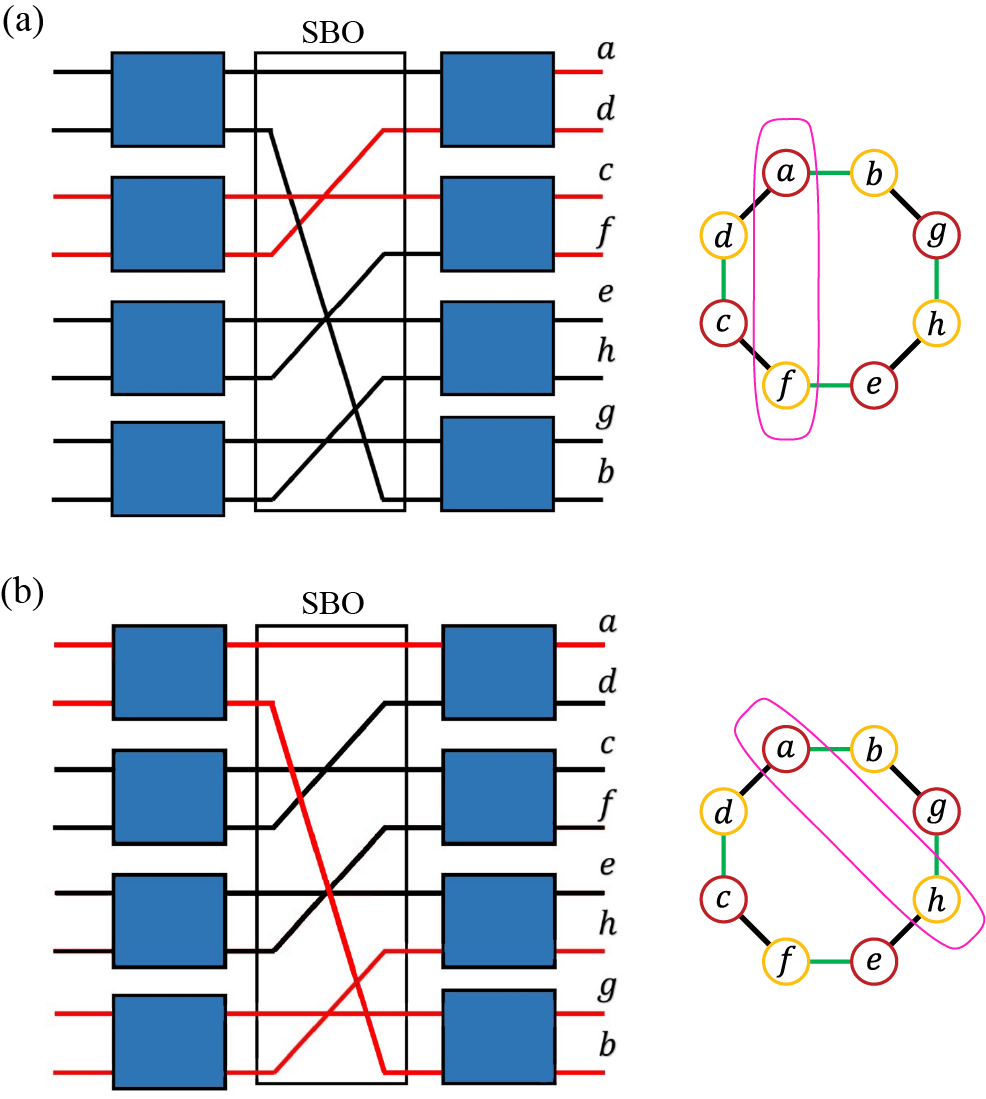}
  \caption{Different cases for bi-partitions of the form 2$\times$6 with one of the partitions composed of third nearest neighbors. (a) Partition composed of modes $a$ and $f$, which are  connected by cascaded PAs. (b) Partition composed of modes $a$ and $h$, which come from completely independent processes. SBO: switchboard operation.}
  \label{twocases}
\end{figure}

Additionally, Fig.~\ref{PPT8-1}(b) shows that the two cases of third nearest neighbors exhibit a different behavior. This can be understood by tracking how the two modes in one of the partitions are connected to each other. The connection shown in Fig.~\ref{twocases}(a) with one partition composed of the third nearest neighbors with two second stage connections and one first stage connection, is a simple cascade PA configuration where the two stages generate a four-mode entangled state. However, when one of the partitions is composed of third nearest neighbors  with two second stage connections and one first stage-connection, as shown in Fig.~\ref{twocases}(b), the modes in the partition come from completely independent processes.

The remaining bi-partitions of the form 3$\times$5, with $\prescript{8\mkern-0.5mu}{}C_{3}=56$ bi-partitions, and 4$\times$4, with $\prescript{8\mkern-0.5mu}{}C_{4}/2=35$ bi-partitions, have smallest symplectic eigenvalues shown in Figs.~\ref{PPT8}(c) and~\ref{PPT8}(d), respectively. Although the number of possible bi-partitions increases rapidly, the physics remain the same. As can be seen from Fig.~\ref{PPT8}, all possible bi-partitions for the octapartite system have  smallest symplectic eigenvalues less than 1 for a squeezing parameter greater than 0. Therefore, the system generates a GME octapartite state.

\begin{table}[t!]
    \centering
    \begin{tabular}{ccccccccccccccc}
        \cellcolor{gray} A & \cellcolor{blue!10} B & \cellcolor{red!25} C &    0 & 0&0 & 0& 0  & 0 &  0 & \dots & 0 & \cellcolor{green!25} D &\cellcolor{red!25} C & \cellcolor{blue!25} B$^{\prime}$ \\  \\[-.5em]
        \cellcolor{blue!10} B & \cellcolor{gray} A & \cellcolor{blue!25} B$^{\prime}$ &  \cellcolor{red!25} C & \cellcolor{green!25} D& 0 & 0&0  & 0 &  0 & \dots & 0 &0 & 0 & \cellcolor{red!25} C \\ \\[-.5em]
        \cellcolor{red!25} C & \cellcolor{blue!25} B$^{\prime}$ & \cellcolor{gray} A & \cellcolor{blue!10} B &  \cellcolor{red!25} C & 0 & 0& 0  & 0 &  0 & \dots & 0 &0 & 0 & \cellcolor{green!25} D \\ \\[-.5em]
        0 & \cellcolor{red!25} C & \cellcolor{blue!10} B & \cellcolor{gray} A & \cellcolor{blue!25} B$^{\prime}$ &  \cellcolor{red!25} C &  \cellcolor{green!25} D &0 & 0&  0 & \dots & 0 &0 & 0 & 0 \\ \\[-.5em]
        0 & \cellcolor{green!25} D & \cellcolor{red!25} C & \cellcolor{blue!25} B$^{\prime}$ & \cellcolor{gray} A &  \cellcolor{blue!10} B &  \cellcolor{red!25} C & 0 & 0& 0 & \dots & 0 &0 & 0 & 0 \\ \\[-.5em]
        0 &0 &0 & \cellcolor{red!25} C & \cellcolor{blue!10} B & \cellcolor{gray} A & \cellcolor{blue!25} B$^{\prime}$ &  \cellcolor{red!25} C &  \cellcolor{green!25} D & 0 & \dots & 0 &0 & 0 & 0 \\ \\[-.5em]
        0 &0 &0 & \cellcolor{green!25} D & \cellcolor{red!25} C & \cellcolor{blue!25} B$^{\prime}$ & \cellcolor{gray} A &  \cellcolor{blue!10} B &  \cellcolor{red!25} C &  0 & \dots & 0 &0 & 0 & 0 \\ \\[-.5em]
        \vdots  &  \vdots  &  \vdots  &  \vdots  &  \vdots  &  \vdots  &  \vdots  &   \vdots  &   \vdots  &   \vdots  & $\ddots$ & 0 &0 & 0 & 0 \\ \\[-.5em]
        0  &  0  &  0  &  0  &   0  &  0  &  \cellcolor{red!25} C  & \cellcolor{blue!10}B & \cellcolor{gray}A &\cellcolor{blue!25}B$^{\prime}$ & \cellcolor{red!25}C & \cellcolor{green!25} D &  0 &  0&  0\\ \\[-.5em]
        0  &  0  &  0  &  0  &   0  &  0  &  \cellcolor{green!25} D  & \cellcolor{red!25}C & \cellcolor{blue!25}B$^{\prime}$ &\cellcolor{gray}A & \cellcolor{blue!10}B & \cellcolor{red!25} C &  0 &  0&  0\\ \\[-.5em]
        0  &  0  &  0  &  0  &  0  &  0  &   0  &  0  &  \cellcolor{red!25} C  & \cellcolor{blue!10}B & \cellcolor{gray}A &\cellcolor{blue!25}B$^{\prime}$ & \cellcolor{red!25}C & \cellcolor{green!25} D &  0 \\ \\[-.5em]
        0  &  0  &  0  &  0  &  0  &  0  &   0  &  0  &  \cellcolor{green!25} D  & \cellcolor{red!25}C & \cellcolor{blue!25}B$^{\prime}$ &\cellcolor{gray}A & \cellcolor{blue!10}B & \cellcolor{red!25} C &  0 \\ \\[-.5em]
        \cellcolor{green!25}D  &  0  &  0  &  0  &  0  &  0  &  0  &   0  &  0  &   0  & \cellcolor{red!25}C & \cellcolor{blue!10}B &\cellcolor{gray}A & \cellcolor{blue!25}B$^{\prime}$ & \cellcolor{red!25} C \\ \\[-.5em]
        \cellcolor{red!25}C  &  0  &  0  &  0  &  0  &  0  &  0  &   0  &  0  &   0  & \cellcolor{green!25}D & \cellcolor{red!25}C &\cellcolor{blue!25}B$^{\prime}$ & \cellcolor{gray}A & \cellcolor{blue!10} B \\ \\[-.5em]
        \cellcolor{blue!25}B$^{\prime}$  &  \cellcolor{red!25}C  &  \cellcolor{green!25}D  &  0  &  0  &  0  &  0  &   0  &  0  &   0  & 0 & 0 &\cellcolor{red!25}C & \cellcolor{blue!10}B & \cellcolor{gray} A \\ \\[-.5em]
    \end{tabular}
    \caption{Matrix structure of the quadrature CM for a $2N$-partite system. Matrix elements labeled as A (grey) represent self-correlations, B (light blue) and B$^{\prime}$ (blue) represent correlations between nearest neighbors, C (red) represent correlations between second nearest neighbors, and D (green) represent correlations between third nearest neighbors.}
    \label{extension}
\end{table}

\begin{figure}[t!]
  \centering
  \includegraphics{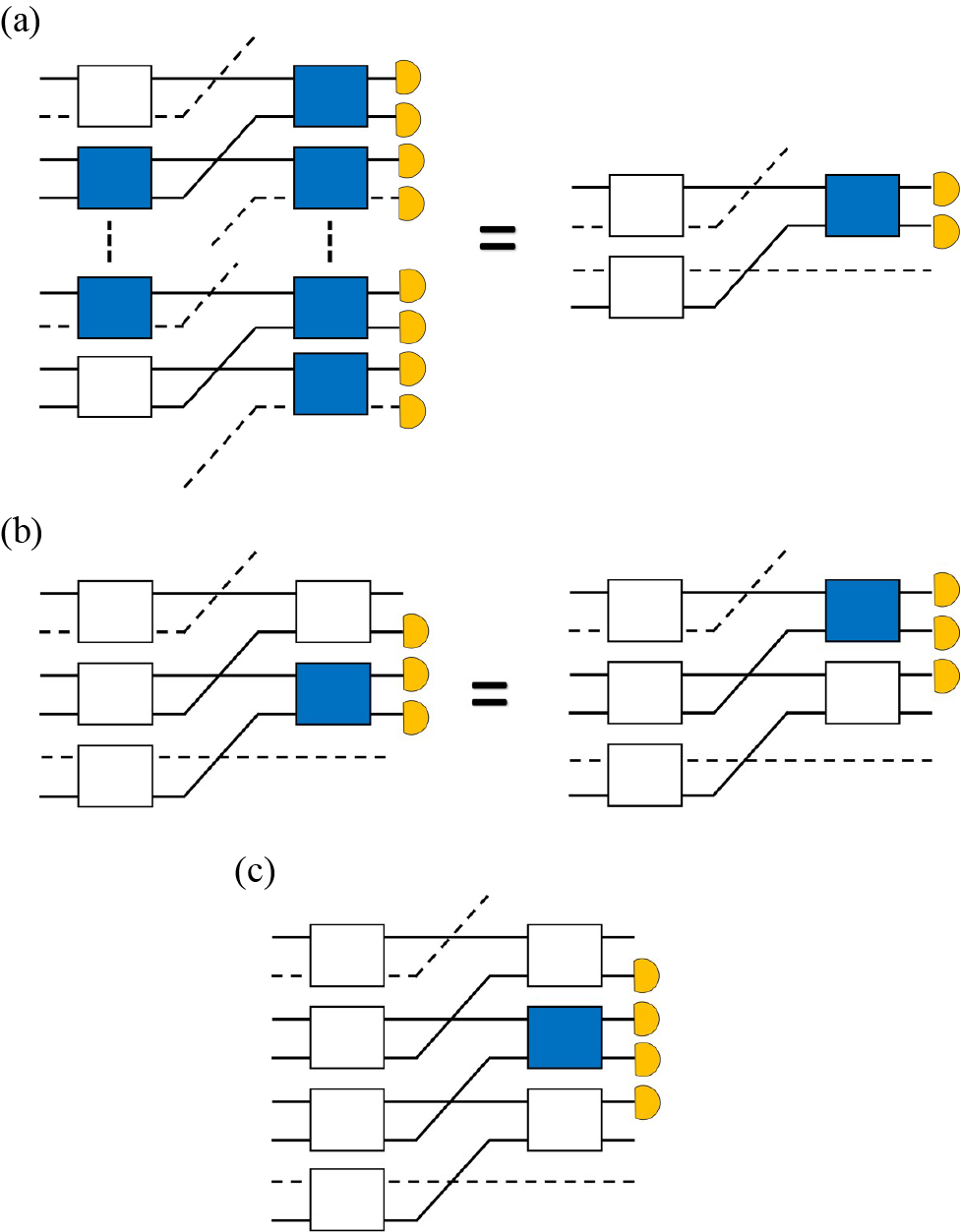}
  \caption{A schematic representation of the unitary reduction that can be implemented. The boxes represent a two-mode squeezing operation, while the yellow semi-circle represents the detected modes (or the modes that are considered part of the one of the bi-partitions or subsystems). We can apply an inverse two-mode squeezing operation on the detected modes to cancel out the two-mode squeezers in blue. Through this reduction: (a) We show that the subsystem on the left-hand side is unitarily equivalent to the one on the right hand side. (b) We show how all odd (3 or more) consecutive modes can be reduced to either of these set-ups. The thermality of these two are equivalent due to the symmetry of the system. (c) We show that all even (4 or more) can be reduced to this set-up, or the set-up shown in (a).}
  \label{Reduction}
\end{figure}

\section{\label{2NM} Generalization to $2N$-Modes}

The results obtained above show that the proposed PA network generates a multipartite quantum state with symmetries that make it possible to understand the correlations between the different modes.  These symmetries become more apparent when we consider the CM. Table~\ref{extension} shows the structure of the CM for an arbitrary number of modes, with martix element that have the same functional form coded with the same color.  The non-zero matrix elements take the form: ${\rm A}=\alpha_{+} \beta_{-} I_{2} $, ${\rm B} = -\alpha_{-} (\beta_{+}+1) J/2$, ${\rm B}^\prime = -\alpha_{+} \beta_{-} J$, ${\rm C} = \alpha_{-} \beta_{-} I_{2}/2$, and ${\rm D} = -\alpha_{-} (\beta_{+}-1) J/2$, where $I_{2}$ and $J$ are the 2$\times$2 identity matrix and the $\textrm{diag}$\{1,-1\} matrix, respectively.  Given that only neighbors with less than a fourth nearest neighbor connection are correlated, as discussed in the Sect.~\ref{HexOct}, most of the elements of the CM are zero with the nonzero ones centered around the diagonal. As can be seen, towards the upper and lower corners of the CM, this leads to some elements wrapping around the edges. Note that the number of modes generated by the proposed PA network is always even, as modes are always added in pairs.

Terms in the CM labeled as A, marked in gray, represent the self-correlation for each mode and correspond to the variances of the quadratures. Since all the modes are amplified equally, they all have the same variance. Terms labeled as B and B$^{\prime}$, marked in light blue and blue, respectively, represent the nearest neighbors. As explained in the previous section, these terms are different given that B results from entangling by a PA from the first stage followed by amplification noise added by the second stage; however, B$^{\prime}$ results from having two modes with amplification noise after the first stage entangled by a PA from the second stage. Finally, terms labeled as C, marked in red, represent second nearest neighbors, and term labeled as D, marked in green, represent third nearest neighbors. Note that some CM matrix elements for third nearest neighbors are zero, which indicates that there are no correlations between the corresponding modes. As shown in Fig.~\ref{twocases}(b), these zero terms corresponds to modes, such as $a$ and $h$, that come from processes that are independent, such that there is not even an indirect "interaction" between them. More details on the matrix elements can be found in App.~\ref{app:A}.

Extending the PPT criteria analysis performed for the hexapartite and octapartite cases to an arbitary  $2N$-partite system to verify the presence of genuine $2N$-partite entanglement would becomes impossible, as the number of bi-partitions that need to be considered increases exponentially. In order to overcome this challenge, we use a measure known as $\alpha$-entanglement of formation and take advantage of the symmetry of the CM to show that some partitions are more entangled than others. The genuine $2N$-mode entanglement entropy for a pure state $\ket{\psi}$ is calculated as follows \cite{SzalayMultipartite,OnoeMultipartite}:
\begin{equation}
    E_{2N}(\ket{\psi})=\underset{P\subset \{1,2,...,2N\}}{\textrm{min}}S[\textrm{Tr}_{P}(\ket{\psi}\bra{\psi})], \label{EQN2NEE}
\end{equation}
where $S[\rho]$ is the Von Neumann entropy defined as $S[\rho]=-\textrm{Tr}(\rho \ln \rho)$ and the minimization is done over all possible subsystems. The pure state $\ket{\psi}$ has genuine $2N$-partite entanglement if $E_{2N}(\ket{\psi})\neq 0$. This is similar to the PPT criterion, whereby the PPT criterion must be violated for all possible bi-partitioning.

Due to the symmetry of the Von Neumann entropy (i.e., $S[\textrm{Tr}_A(\rho_{AB})]=S[\textrm{Tr}_B(\rho_{AB})]$)~\cite{WeedbrookGQI}, it is sufficient to conduct the minimization over the cases where the number of modes in $P\subset \{1,2,...,2N\}$  is less than $N$. For Gaussian states, the Von Neumann entropy of an $M$-mode mixed state $\rho_M$ can be computed via its CM, $\sigma_{M}$, according to~\cite{WeedbrookGQI}
\begin{gather}
    S(\sigma_{M})=\frac{1}{2}\sum_{n=1}^M h(\nu_n),\\
    h(x)=\frac{x_+}{2}\log_2\left(\frac{x_+}{2}\right)-\frac{x_-}{2}\log_2\left(\frac{x_-}{2}\right),
\end{gather}
where $\nu_n$ is the $n^{\text{th}}$ symplectic eigenvalue of $\sigma_M$ and $x_{\pm}=x\pm 1$.

\begin{figure}[t]
  \centering
  \includegraphics{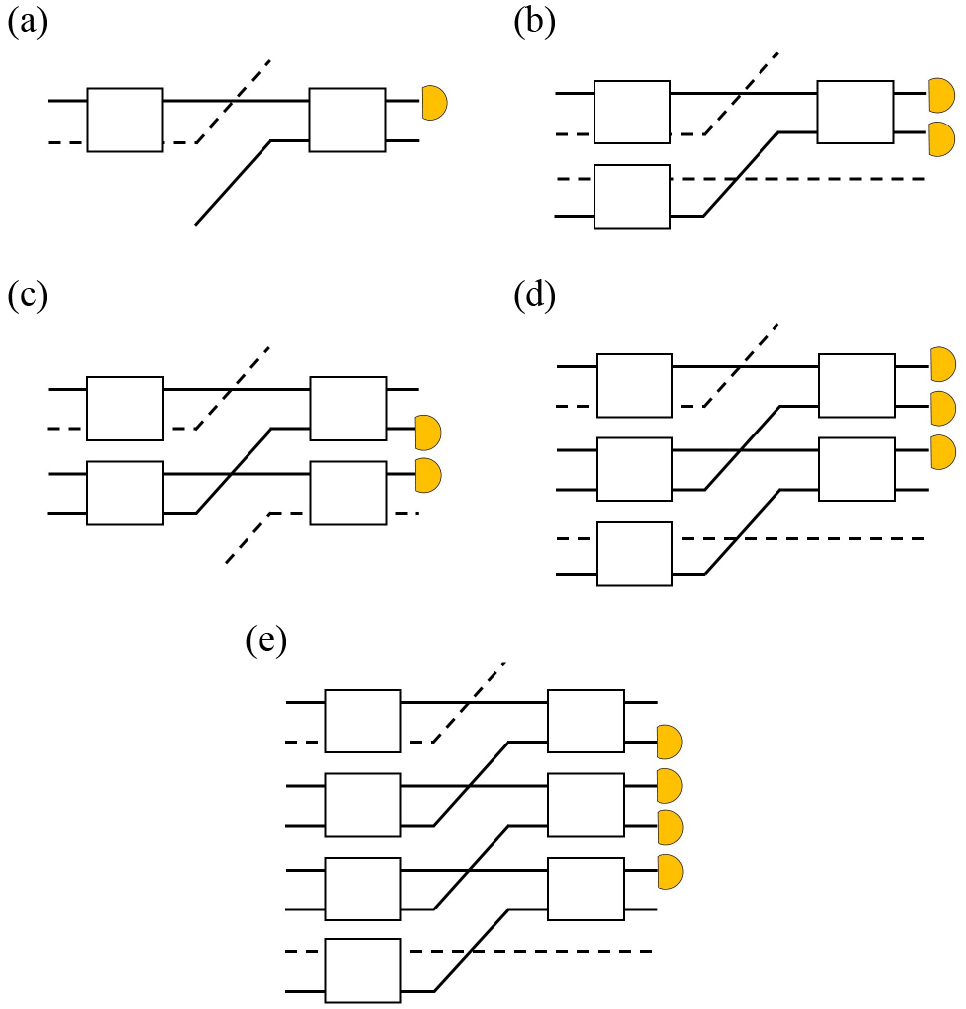}
  \caption{Subsystems that minimizes the entanglement entropy (for more detail see App. \ref{app:D}). The optimization of Eq.~(\ref{EQN2NEE}) needed for the $\alpha$-entanglement of formation can be done by numerically calculating the entanglement entropy when the subsystem is give by one of the 5 cases shown in (a) through (e).}
  \label{Minimum}
\end{figure}

For the CM shown in Table~\ref{extension}, the correlations are strongest between nearest neighbors. As a result, the Von-Neumann entropy of the reduced state $\rho_{P_1}$ is minimized when the subsystem is composed of consecutive modes, e.g. when $P_1=\{1,2,3,4\}$ (see App.~\ref{app:D} for more details). Furthermore, we can unitarily reduce the number of consecutive modes by utilizing the following property:
\begin{equation}
    S(\hat{U}_{P_1}^{\dag}\rho_{P_1}\hat{U}_{P_1})=S(\rho_{P_1}),
\end{equation}
where $\hat{U}_{P_1}$ is an arbitrary unitary operation on the $P_1$ subsystem. By taking advantage of this property, it can be shown that the entropy of an arbitrary number of consecutive modes can be reduced down to the entropy of only two, three, or four modes. This reduction is shown in Fig.~\ref{Reduction}, and makes it possible to extend the analysis to an arbitrary number of $2N$ modes. As an example, we mathematically show this result for a particular case:
\begin{align}
    S(\sigma_{1234})&=S(\hat{U}_{1234}^{\dag}\sigma_{1234}\hat{U}_{1234} )\\
    &=S(\sigma_1 \otimes \mathds{1}_2\otimes\mathds{1}_3\otimes\sigma_4)\\
    &=S(\sigma_1 \otimes \sigma_4)
    \label{EQNreduction}
\end{align}
where $\hat{U}_{1234}=\hat{S}_{12}^{-1}\hat{S}_{34}^{-1}\hat{S}_{23}^{-1}$, with $\hat{S}_{ij}$ representing the two-mode squeezing operation between modes $i$ and $j$. Following the result of App.~\ref{app:D} and this reduction, it can be shown that the Von Neumann entropy for all possible bi-partitioning is larger than or equal to at least one of the five possible partitionings shown in Fig.~\ref{Minimum}. Thus, it is sufficient to verify that the minimum entropy of these five possible partitionings is greater than zero to show GME in the $2N$-mode state. Their entropies are a function of $s_1$ and $s_2$, meaning that the genuine $2N$-partite entanglement entropy stays constant for all $N>3$.

\begin{figure}[t!]
  \centering
  \includegraphics{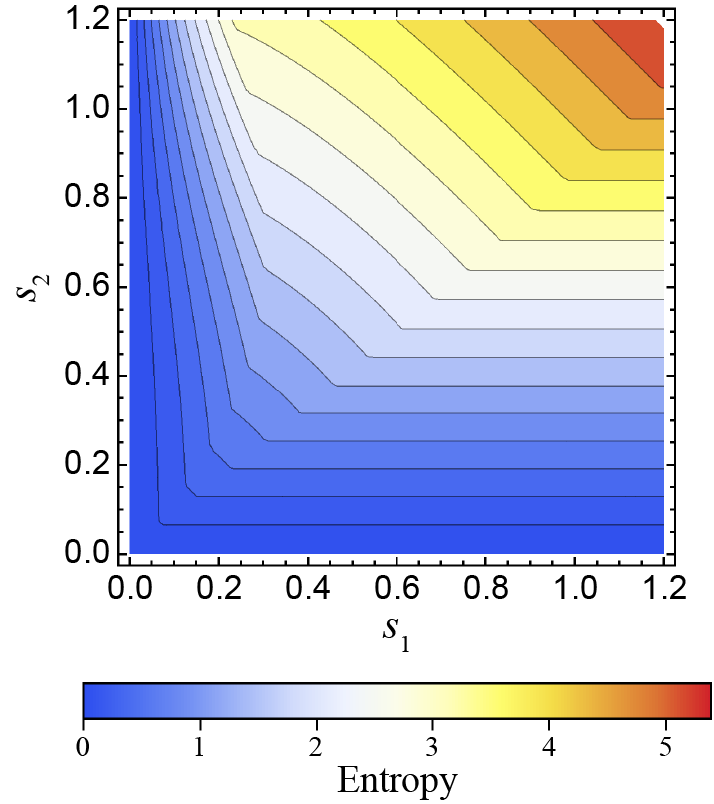}
  \caption{ Minimum entropy of $\{$1$\}$, $\{$1,2$\}$, $\{$2,3$\}$, $\{$1,2,3$\}$, $\{$2,3,4$\}$, $\{$1,2,3,4$\}$, $\{$2,3,4,5$\}$, etc. We plot the $2N$-mode entanglement entropy as a function of $s_1$ and $s_2$.}
  \label{FigPlotMain}
\end{figure}

We plot the genuine $2N$-partite entanglement entropy in Fig. \ref{FigPlotMain}. As can be seen, the miniumum entropy for the system is greater than zero for any value of $s_{1}>0$ and $s_{2}>0$, which shows that the generated state contains genuine $2N$-partite entanglement. Note that the simplicity of the reduction method, and the constant entanglement of entropy for $N$-modes, for $N>3$, can be attributed to the symmetric set-up of the PA network. As a result, it is possible to show theoretically that the proposed scheme is scalable and can generate genuine $2N$-partite entanglement for arbitrarily large $N$ through careful selection of the switchboard operation, as shown in Fig.~\ref{multibeam}(c).

\section{Conclusion}

In conclusion, we propose a novel technique for generating scalable GME using a PA network. Our proposed scheme offers an efficient alternative for generating genuine multipartite entangled states that can scale to a larger number of parties due to the symmetries in the correlations between the different modes. We show the presence of quadripartite, hexapartite, and octapartite GME through a violation of the PPT criteria for all possible bi-partitions.  The symmetry in the system make it possible to track the correlations between all the modes and explain the physical mechanisms behind them.

We show that it is possible to generalize the covariance matrix to a $2N$-partite system with a high degree of symmetry in which correlations are present between third nearest neighbors at most. Such highly symmetric covariance matrix makes it possible to verify the presence of genuine $2N$-partite entanglement through the use $\alpha$-entanglement of formation by taking advantage of these symmetries to reduce all possible bi-partitons to only  5 cases. By showing that the Von Neumann entropy for these 5 bi-partitions is larger than zero, we are able to show that the proposed scheme generates a scalable genuine $2N$-partite entangled state for aribtrarily large $N$.

The two-stage design of the parametric network, in combination with the proposed swithboard operation, make the proposed scheme amenable to an implementation that takes advantage of the multispatial mode properties of parametric amplifiers~\cite{Boyer08,ashok,PhysRevLett.124.090501}. Such an approach would lead to a highly scalable and compact source of GME. Furthermore, the proposed technique can be extended to generate quantum states with more complex connections, such as ones with a cubic graphical representation, through the addition of additional PA stages to the network.

\section*{Acknowledgements}
This work was supported by the National Science Foundation (NSF grant PHYS-1752938). AMM acknowledges support from the US Department of Energy, Office of Science, National Quantum Information Science Research Centers, Quantum Science Center for finalizing the manuscript.

\appendix

\section{\label{app:A}Covariance Matrices for the Quadripartite, Hexapartite, and Octapartite Systems}

The CM for the quadrapartite system is an 8$\times$8 matrix with two components, one for each quadrature, per mode. Taking the definition of the CM as $\sigma=<\bm{\xi}\bm{\xi}^T>$, with $\bm{\xi}=[\hat{X}_a,\hat{Y}_a,\hat{X}_b,\hat{Y}_b,\hat{X}_c,\hat{Y}_c,\hat{X}_d,\hat{Y}_d]$ for the four mode system, and the two-mode squeezing transformation $\hat{S}_{ij} (\zeta)=\textrm{exp}(\zeta^* \hat{i}\hat{j}-\zeta \hat{i}^\dagger \hat{j}^\dagger)$ between input modes $i$ and $j$ performed by each PA, the matrix elements of the CM in terms of the quadratures of the four modes takes the form
\begin{widetext}
    \begin{eqnarray}
        \sigma_{4}&=&\left(
        \begin{array}{cccccccc}
            \left\langle \hat{X}_a^2\right\rangle  & 0 & \left\langle \hat{X}_a \hat{X}_b\right\rangle  & 0 & \left\langle \hat{X}_a \hat{X}_c\right\rangle  & 0 & \left\langle \hat{X}_a \hat{X}_d\right\rangle  & 0 \\
            0 & \left\langle \hat{Y}_a^2\right\rangle  & 0 & \left\langle \hat{Y}_a \hat{Y}_b\right\rangle  & 0 & \left\langle \hat{Y}_a \hat{Y}_c\right\rangle  & 0 & \left\langle \hat{Y}_a \hat{Y}_d\right\rangle  \\
            \left\langle \hat{X}_a \hat{X}_b\right\rangle  & 0 & \left\langle \hat{X}_b^2\right\rangle  & 0 & \left\langle \hat{X}_b \hat{X}_c\right\rangle  & 0 & \left\langle \hat{X}_b \hat{X}_d\right\rangle  & 0 \\
            0 & \left\langle \hat{Y}_a \hat{Y}_b\right\rangle  & 0 & \left\langle \hat{Y}_b^2\right\rangle  & 0 & \left\langle \hat{Y}_b \hat{Y}_c\right\rangle  & 0 & \left\langle \hat{Y}_b \hat{Y}_d\right\rangle  \\
            \left\langle \hat{X}_a \hat{X}_c\right\rangle  & 0 & \left\langle \hat{X}_b \hat{X}_c\right\rangle  & 0 & \left\langle \hat{X}_c^2\right\rangle  & 0 & \left\langle \hat{X}_c \hat{X}_d\right\rangle  & 0 \\
            0 & \left\langle \hat{Y}_a \hat{Y}_c\right\rangle  & 0 & \left\langle \hat{Y}_b \hat{Y}_c\right\rangle  & 0 & \left\langle \hat{Y}_c^2\right\rangle  & 0 & \left\langle \hat{Y}_c \hat{Y}_d\right\rangle  \\
            \left\langle \hat{X}_a \hat{X}_d\right\rangle  & 0 & \left\langle \hat{X}_b \hat{X}_d\right\rangle  & 0 & \left\langle \hat{X}_c \hat{X}_d\right\rangle  & 0 & \left\langle \hat{X}_d^2\right\rangle  & 0 \\
            0 & \left\langle \hat{Y}_a \hat{Y}_d\right\rangle  & 0 & \left\langle \hat{Y}_b \hat{Y}_d\right\rangle  & 0 & \left\langle \hat{Y}_c \hat{Y}_d\right\rangle  & 0 & \left\langle \hat{Y}_d^2\right\rangle  \\
        \end{array}
        \right)\nonumber\\
        &\equiv&\left(
        \begin{array}{cccccccc}
            \alpha _+ \beta _+ & 0 & \alpha _- \beta _+ & 0 & \alpha _- \beta _- & 0 & \beta _- \alpha _+ & 0 \\
            0 & \alpha _+ \beta _+ & 0 & -\alpha _- \beta _+ & 0 & \alpha _- \beta _- & 0 & -\beta _- \alpha _+ \\
            \alpha _- \beta _+ & 0 & \alpha _+ \beta _+ & 0 & \beta _- \alpha _+ & 0 & \alpha _- \beta _- & 0 \\
            0 & -\alpha _- \beta _+ & 0 & \alpha _+ \beta _+ & 0 & -\beta _- \alpha _+ & 0 & \alpha _- \beta _- \\
            \alpha _- \beta _- & 0 & \beta _- \alpha _+ & 0 & \alpha _+ \beta _+ & 0 & \alpha _- \beta _+ & 0 \\
            0 & \alpha _- \beta _- & 0 & -\beta _- \alpha _+ & 0 & \alpha _+ \beta _+ & 0 & -\alpha _- \beta _+ \\
            \beta _- \alpha _+ & 0 & \alpha _- \beta _- & 0 & \alpha _- \beta _+ & 0 & \alpha _+ \beta _+ & 0 \\
            0 & -\beta _- \alpha _+ & 0 & \alpha _- \beta _- & 0 & -\alpha _- \beta _+ & 0 & \alpha _+ \beta _+ \\
        \end{array}
        \right).
        \label{a1}
    \end{eqnarray}
\end{widetext}
where $\alpha _{\pm} = (e^{-2 s_1}\pm e^{2 s_1})/2$, and $\beta _{\pm}= (e^{-2 s_2}\pm e^{2 s_2})/2$. Note that the two-mode squeezing transformation does not lead to correlations between the amplitude and phase quadratures of the modes, which results in the zero matrix elements of the CM.

When we extend the parametric amplifier network to a hexapartite system, the CM becomes a 12$\times$12 matrix, with a total of 12 quadratures to consider. Expanding on the results of the quadrapartite case and using the notation introduced in Sect.~\ref{2NM}, the CM takes the form
\begin{equation}
   \sigma_{6}=
   \begin{tabular}{cccccc}
        \cellcolor{gray} A & \cellcolor{blue!10} B & \cellcolor{red!25} C & \cellcolor{green!25} D &\cellcolor{red!25} C & \cellcolor{blue!25} B$^{\prime}$ \\  \\[-.5em]
        \cellcolor{blue!10} B & \cellcolor{gray} A & \cellcolor{blue!25} B$^{\prime}$ &  \cellcolor{red!25} C & \cellcolor{green!25} D& \cellcolor{red!25} C \\ \\[-.5em]
        \cellcolor{red!25} C & \cellcolor{blue!25} B$^{\prime}$ & \cellcolor{gray} A & \cellcolor{blue!10} B &  \cellcolor{red!25} C & \cellcolor{green!25} D \\ \\[-.5em]
        \cellcolor{green!25}D  &  \cellcolor{red!25}C & \cellcolor{blue!10}B &\cellcolor{gray}A & \cellcolor{blue!25}B$^{\prime}$ & \cellcolor{red!25} C \\ \\[-.5em]
        \cellcolor{red!25}C  & \cellcolor{green!25}D & \cellcolor{red!25}C &\cellcolor{blue!25}B$^{\prime}$ & \cellcolor{gray}A & \cellcolor{blue!10} B \\ \\[-.5em]
        \cellcolor{blue!25}B$^{\prime}$  &  \cellcolor{red!25}C  &  \cellcolor{green!25}D  &\cellcolor{red!25}C & \cellcolor{blue!10}B & \cellcolor{gray} A
    \end{tabular},
\end{equation}
where each of the blocks is a two dimensional submatrix composed of an amplitude and an a phase quadrature for the different modes. Simarly, the results can be extended to the case of eight modes, for the which the CM is a 16$\times$16 matrix of the form
\begin{equation}
    \sigma_{8}=
    \begin{tabular}{cccccccc}
        \cellcolor{gray} A & \cellcolor{blue!10} B & \cellcolor{red!25} C &0&0& \cellcolor{green!25} D &\cellcolor{red!25} C & \cellcolor{blue!25} B$^{\prime}$ \\  \\[-.5em]
        \cellcolor{blue!10} B & \cellcolor{gray} A & \cellcolor{blue!25} B$^{\prime}$ &  \cellcolor{red!25} C & \cellcolor{green!25} D&0&0& \cellcolor{red!25} C \\ \\[-.5em]
        \cellcolor{red!25} C & \cellcolor{blue!25} B$^{\prime}$ & \cellcolor{gray} A & \cellcolor{blue!10} B &  \cellcolor{red!25} C &  0&0&\cellcolor{green!25}D \\ \\[-.5em]
        0&\cellcolor{red!25} C & \cellcolor{blue!25} B$^{\prime}$ & \cellcolor{gray} A & \cellcolor{blue!10} B &  \cellcolor{red!25} C &\cellcolor{green!25}D &0\\ \\[-.5em]
        0 & \cellcolor{green!25}D & \cellcolor{red!25}C &\cellcolor{blue!25}B$^{\prime}$ & \cellcolor{gray}A & \cellcolor{blue!10} B&\cellcolor{red!25}C&0 \\ \\[-.5em]
        \cellcolor{green!25}D&0&0  &  \cellcolor{red!25}C & \cellcolor{blue!10}B &\cellcolor{gray}A & \cellcolor{blue!25}B$^{\prime}$ & \cellcolor{red!25} C \\ \\[-.5em]
        \cellcolor{red!25}C &0&0 & \cellcolor{green!25}D & \cellcolor{red!25}C &\cellcolor{blue!25}B$^{\prime}$ & \cellcolor{gray}A & \cellcolor{blue!10} B \\ \\[-.5em]
        \cellcolor{blue!25}B$^{\prime}$  &  \cellcolor{red!25}C  &  \cellcolor{green!25}D&0&0  &\cellcolor{red!25}C & \cellcolor{blue!10}B & \cellcolor{gray} A
    \end{tabular},
\end{equation}
where the presence of connection higher to third nearest neighbor start to appear and the general structure of the CM as the system scales to a large number of modes becomes evident.

\begin{figure*}[t]
    \centering
    \includegraphics{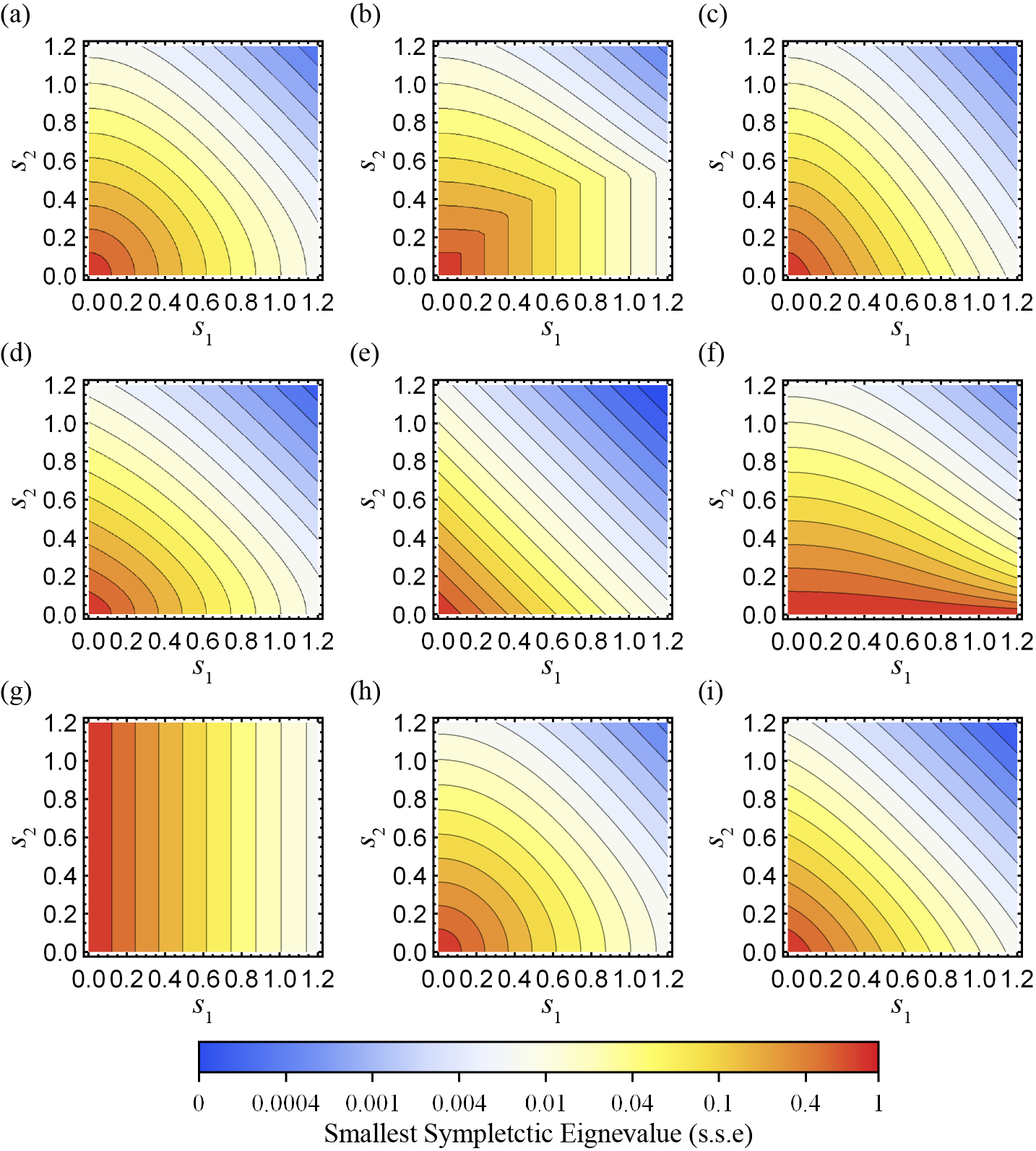}
    \caption{PPT criteria for the cases of bi-partitions of the form (a) 1$\times$5 for partitions ($a|bcdef$, $b|acdef$, $c|abdef$, $d|abcef$, $e|abdf$, $f|abcde$), (b) 3$\times$3 for partitions ($abe|cdf$, $bef|acd$, $cef|abd$), (c) 3$\times$3 for partitions ($ade|bcf$, $adf|bce$, $bde|acf$), (d)  3$\times$3 for partitions ($abc|def$, $abf|cde$, $aef|bcd$), (e) 3$\times$3 for partitions ($ace|bdf$, $bdf|ace$), (f) 2$\times$4 for partitions ($ab|cdef$, $ef|abcd$, $cd|abef$), (g) 2$\times$4 for partitions ($ad|bcef$, $be|acdf$, $cf|abde$), (h) 2$\times$4 for partitions ($af|bcde$, $bc|adef$, $de|abcf$), and (i) 2$\times$4 for partitions ($ae|bcdf$, $bf|acde$, $ce|abdf$, $df|abce$, $ac|bdef$, $bd|acef$).}
    \label{PPT2appen1}
\end{figure*}

\section{\label{app:B}PPT Criteria for Hexapartite System}

In the main text we limited the discussion of the PPT criteria to the case in which both parametric stages have the same level of squeezing, $s_{1}=s_{2}$.  Here, we show the smallest symplectic eigenvalues for all possible bi-partitions as a function of the squeezing parameters of the first and the second stages. Figure~\ref{PPT2appen1}(a) shows bi-partitions of form 1$\times$5 ($a|bcdef$, $b|acdef$, $c|abdef$, $d|abcef$, $e|abdf$, $f|abcde$). As described in the main text, due to symmetry they all exhibit the same behavior.

Figures~\ref{PPT2appen1}(b) through~\ref{PPT2appen1}(e) show the four cases of bi-partitions of form 3$\times$3 with: one of the partitions composed only of nearest neighbors ($abe|cdf$, $bef|acd$, $cef|abd$), one of the partitions that group nearest neighbors connected through a second stage and a second nearest neighbor mode ($ade|bcf$, $adf|bce$, $bde|acf$), one of the partitions that group nearest neighbors connected through a first stage and a second nearest neighbor mode ($abc|def$, $abf|cde$, $aef|bcd$), and one of the partitions composed only of second nearest neighbors ($ace|bdf$, $bdf|ace$), respectively.  Note that in Fig.~\ref{PPT2appen1}(b) the plot exhibits a discontinuous behavior due to the fact that we are choosing the smallest symplectic eigenvalue and there is a competition between PAs from the first stage and the second stage.

Figures~\ref{PPT2appen1}(f) through~\ref{PPT2appen1}(i) show four cases of bi-partitions of the form 2$\times$4: bi-partitions between nearest neighbors connected through a first stage and the rest of the system ($ab|cdef$, $ef|abcd$, $cd|abef$), bi-partitions between nearest neighbors connected with a second stage and the rest of the system ($ad|bcef$, $be|acdf$, $cf|abde$), bi-partitions between third nearest neighbors and the rest of the system ($af|bcde$, $bc|adef$, $de|abcf$), and bi-partitions between second nearest neighbors and the rest of the system ($ae|bcdf$, $bf|acde$, $ce|abdf$, $df|abce$, $ac|bdef$, $bd|acef$).

As was the case for equal level of squeezing in both PA stages,  the smallest symplectic eigenvalues for all possible bi-partitions are always less than 1 when $s_1>0$ and $s_2>0$, which implies a violation of the PPT criteria for all cases.  This shows that even in the general case of different squeezing parameters, the system always generates genuine hexapartite entanglement.

\section{Two Distinct Cases for Hexapartite System}\label{A:app4}

\begin{figure}
    \centering
    \includegraphics{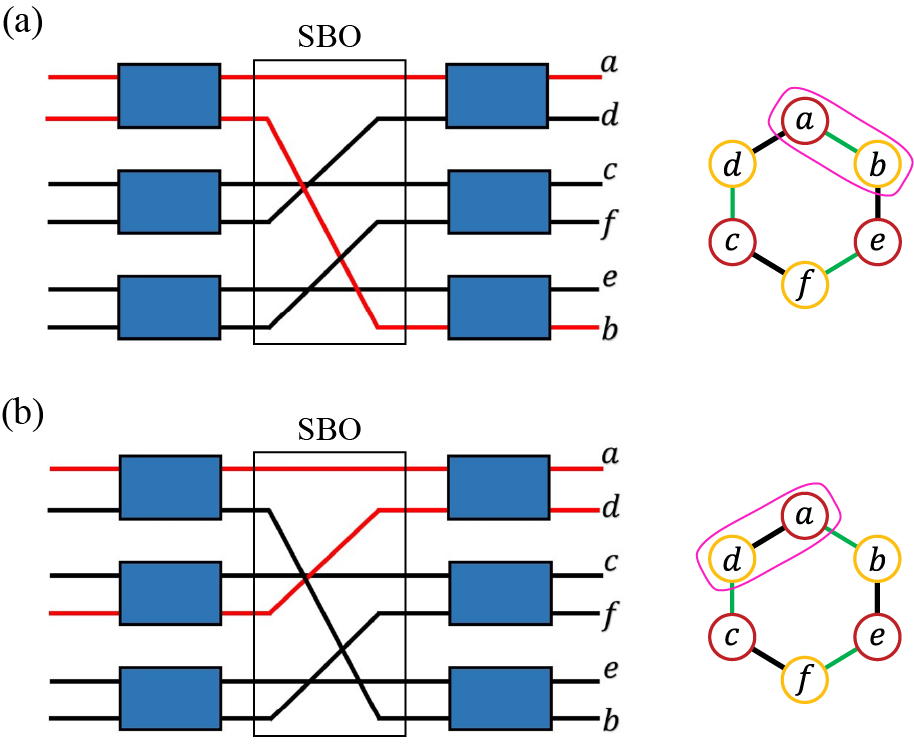}
    \caption{The two possible cases shown in~Fig.~\ref{PPT6-1&2}(c) for bi-partitions of the form 2$\times$4 with one of the partitions composed of nearest neighbor (a) $a$ and $b$ connected by a first stage PA and (b) $a$ and $d$ connected by a second stage PA.}
    \label{twocasesApp}
\end{figure}

As explained in Sect.~\ref{HexOct}, for the hexapartite system we find that bi-partitions of the form 2$\times$4 with one of the partitions composed of nearest neighbors with a first stage connection show more entanglement than ones with a second stage connection. As we have discussed in Sect.~\ref{HexOct}, the ones with a first stage connection have an entangling operation applied first followed by the addition of amplification noise, while the ones with a second stage connection have the amplification noise added first followed by an entangling operation, as shown in Fig.~\ref{twocasesApp}.
    
We calculate the variance of the photon number difference between the two relevant modes for the cases in which the entangling operation occurs first and the one for which amplification noise is added first and find that they take the form,
\begin{widetext}
    \begin{equation}
        \begin{aligned}
            V_{ab}&=2 \sinh ^2(\text{$s_2$}) \big[ \cosh ^4(\text{$s_1$}) \cosh ^2(\text{$s_2$})+\sinh ^4(\text{$s_1$}) \cosh ^2(\text{$s_2$})+\sinh ^2(\text{$s_1$}) \cosh ^2(\text{$s_1$}) \sinh^2(\text{$s_2$})\big], \\
            V_{ad}&=2 \sinh ^2(\text{$s_1$}) \cosh ^2(\text{$s_1$}),
        \end{aligned}
    \end{equation}
\end{widetext}
where the variance is defined as $V_{ij}=<\Delta(\hat{n}_i-\hat{n}_j)^2>$ and $\hat{n}$ is the number operator. While the variance between the two modes that are entangled in the first stage shows a dependence on the squeezing parameters of the first and the second stages, the variance between the two modes that are entangled in the second stage only shows a dependence on the squeezing parameter of the first stage.

For the case in which we assume the same squeezing parameter for the first and second PA stages, $s_1=s_2\equiv s$, the ratio between the two variance is given by $V_{ab}$/$V_{ad}=2 \sinh ^4(s)+\cosh ^4(s)$. This ratio increases exponentially as the squeezing parameter increase, which shows that breaking the connection introduced by the second stage, see Fig.~\ref{twocasesApp}(a), leads to more excess noise than breaking a connection introduced by the first stage, see Fig.~\ref{twocasesApp}(b). This is why we find a stronger violation of the PPT criteria, and thus a higher degree of entanglement, for bi-partitions of the form shown on the left of Fig.~\ref{PPT6-1&2}(c).

\begin{figure*}
  \centering
  \includegraphics{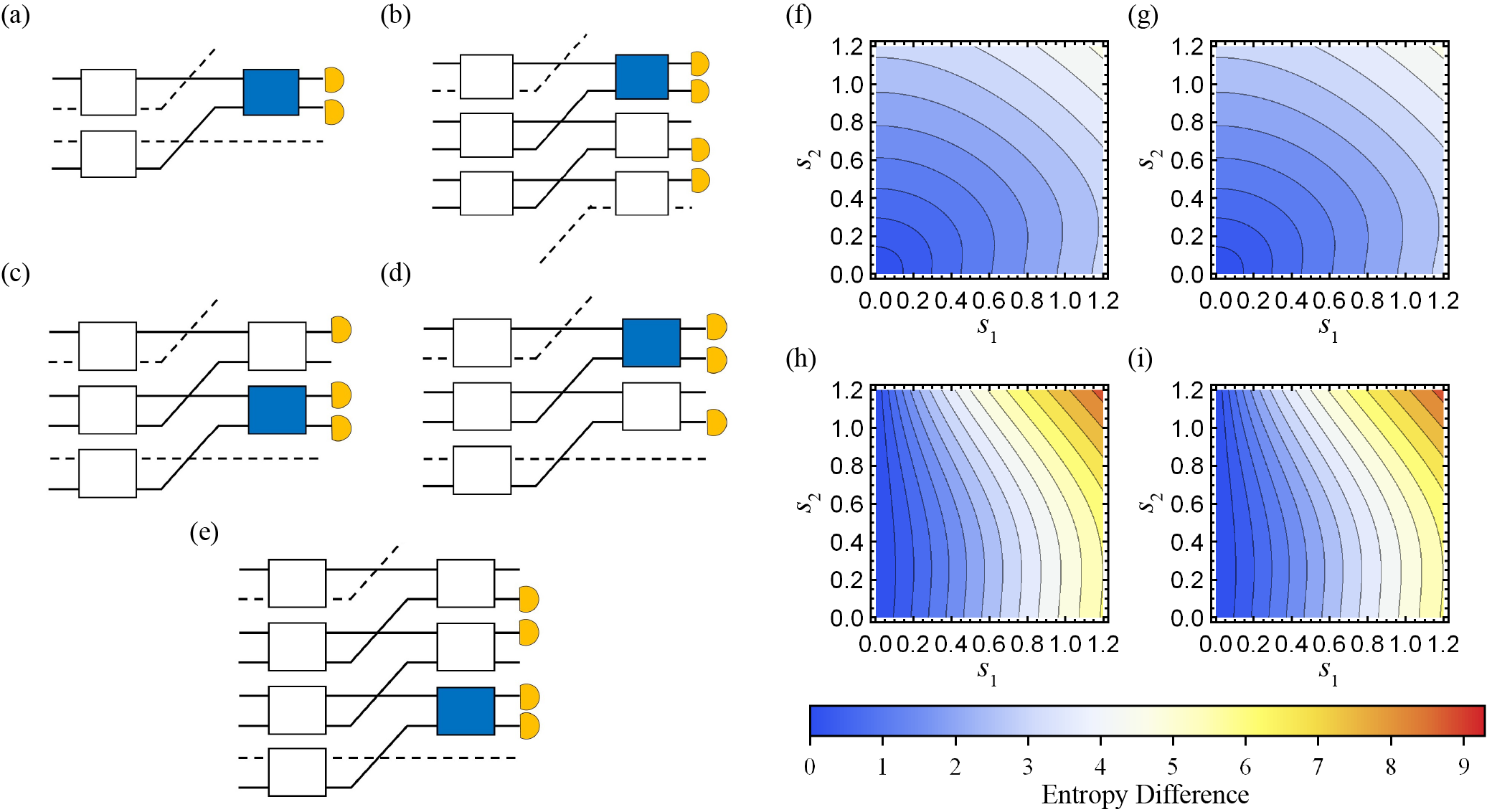}
  \caption{This figure is split into two parts, (a) through (e) and (f) through (i). (b) through (e) show the (disjointed) partitions that can be extended from (a), without an analytical proof that the entropy is equal to or larger than  the partitions shown in Fig.~\ref{Minimum}. We show the entropy associated with the partitions of (b) through (e) are indeed larger than (a) via (f) through (i). These plot show the entropy difference  $S(\textrm{Tr}_{A}(\sigma))-S(\textrm{Tr}_{B}(\sigma))$ where $A$ and $B$ are (f) through (i): $A=\{3,5,6\}$, $\{3,5,7\}$, $\{2,3,5,6\}$, $ \{5,6,8,9\}$ for $B=\{5,6\}$}
 \label{reduction1}
\end{figure*}

\section{\label{app:D} Genuine $2N$-Partite Entanglement}

\begin{figure*}
  \centering
  \includegraphics{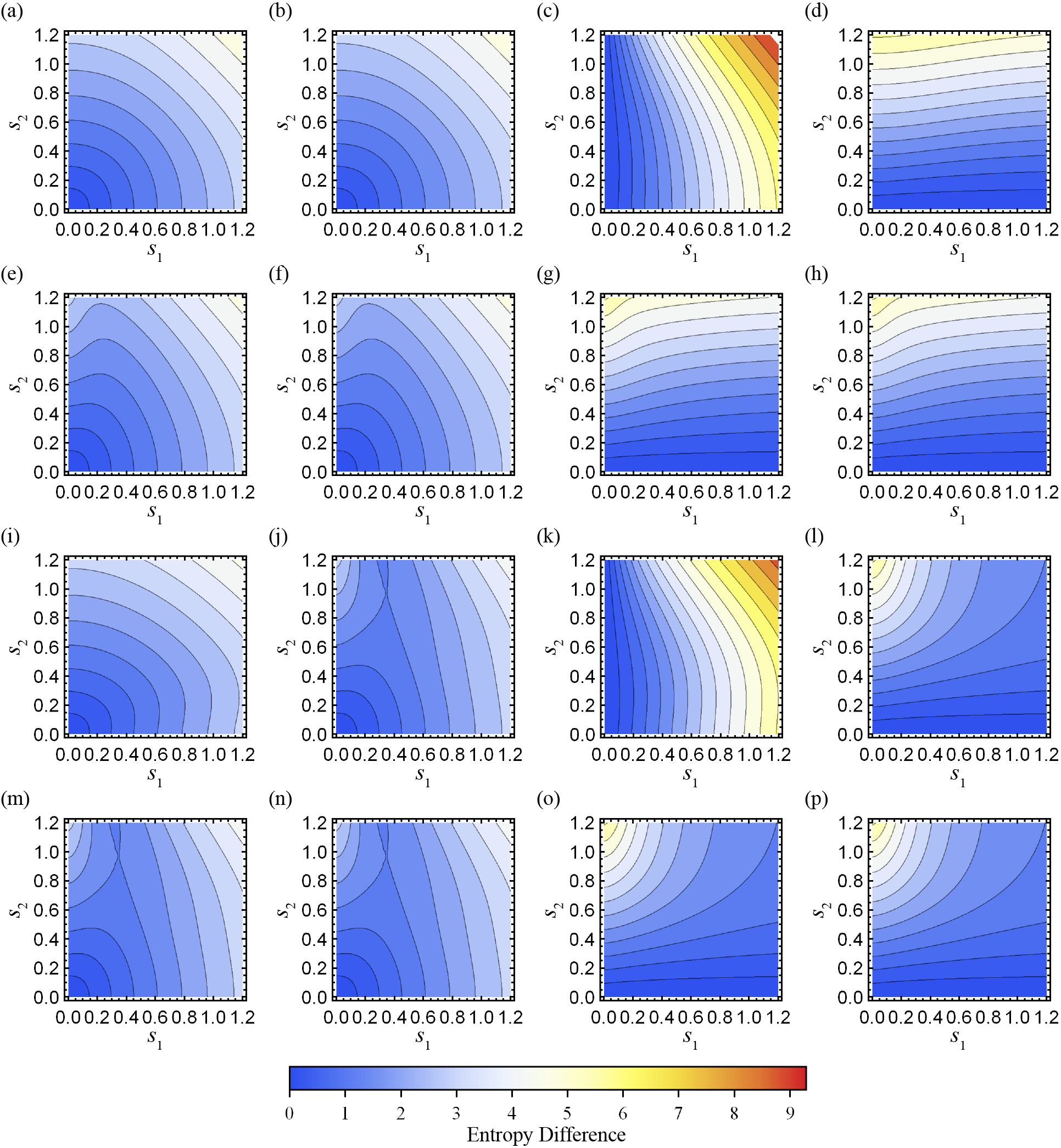}
  \caption{Entropy difference $S(\textrm{Tr}_{A}(\sigma))-S(\textrm{Tr}_{B}(\sigma))$ with $A$ and $B$ given by (a) through (d): $A=\{3,5\}$, $\{5,7\}$, $\{2,3,5\}$, $\{5,7,8\}$ for $B=\{5\}$, (e) through (h) $A=\{4,6,7\}$, $\{6,7,9\}$, $\{3,4,6,7\}$, $\{6,7,9,10\}$ for $B=\{6,7\}$, (i) through (l) $A=\{3,5,6,7\}$, $\{5,6,7,9\}$, $\{2,3,5,6,7\}$, $\{5,6,7,9,10\}$ for $B=\{5,6,7\}$, and (m) through (p)   $A=\{2,4,5,6,7\}$, $\{4,5,6,7,9\}$, $\{1,2,4,5,6,7\}$, $\{4,5,6,7,9,10\}$ for $B=\{4,5,6,7\}$}
  \label{EntropyDiff1}
\end{figure*}

We now show that the entanglement entropy is minimized for subsystems that consists of consecutive modes (i.e. when $P_C=\{M,M+1,M+2,...,M+M'\}$, where $M,M' \in \mathbb{Z}^+$). As an example $P_1=\{1,2,3,4\}$ consists of consecutive modes, while $P_1=\{1,2,3,5\}$ does not.

The first observation is that the correlations only exist between modes that are, at most, three modes apart from each other. This means that when the subsystem consists of groups of modes that are four modes apart from each other, then we can write $\rho_{\{M,...,M+M', M+M'+4,...\}}=\rho_{\{M,...,M+M'\}}\otimes\rho_{\{M+M'+4,...\}}$. As a result, we have that $S(\rho_{\{M,...,M+M',M+M'+4,...\}})=S(\rho_{\{M,...,M+M'\}})+S(\rho_{\{M+M'+4,...\}})$.

The second observation that is made, is that matrices B, B', and D are all proportional to $\textrm{diag}$\{1,-1\}, which corresponds to the squeezing correlations between the two modes (i.e. quantum correlations arising from $\braket{\hat{a}\hat{b}}$ and $\braket{\hat{a}^{\dag}\hat{b}^{\dag}}$ terms). From a simple calculation, we find that the values of B and B' are larger than those of D for all possible squeezing parameters, that is
\begin{widetext}
    \begin{align}
        \sqrt{|B|}-\sqrt{|D|}&=\sinh (2 s_1) \geq 0,\\
        \sqrt{|B^{\prime}|}-\sqrt{|D|}&=\frac{1}{2} \sinh (s_2) [3 \cosh (2 s_1-s_2)+\cosh (2 s_1+s_2)]\geq0.
    \end{align}
\end{widetext}
This implies that the entropy of a state $P_1=p_1\oplus p_2$ is always smaller than $P_1=p_1\oplus (p_2+\vec{2})$, where $p_2=\{N_1,N_2,N_3,...\}$ and $p_2+\vec{2}=\{N_1+2,N_2+2,N_3+2,...\}$. As a result, adding ``islands'', by which we mean groups of consecutive modes, that are more than 2 modes apart from the partitioning will always increase the entropy.

The third observation is that the whole state $\rho=\ket{\psi}\bra{\psi}$ is a pure state. The entropy of the subsystem $\rho_{P_C}$ thus comes from the entanglement between subsystem $P_C$ and the rest of the system. From Table~\ref{extension} in Sect.~\ref{2NM} we can see that correlations between subsystem $P_C$ and the rest of the system only exists for the edge entries $M$, $M+1$, $M+2$, $M+M'-2$, $M+M'-1$ and $M+M'$. This means that tracing out any other modes can only increase the entropy, as it does not remove the entanglement between $\rho_{P_C}$ and the rest of the system. For example, adding consecutive modes will result in an entropy less than adding ``islands'' that are more than 2 modes apart.

The fourth observation is that matrix C of the CM is proportional to $\textrm{diag}\{1,1\}$, which corresponds to classical correlation (i.e. photon counting correlations arising from $\braket{\hat{a}\hat{b}^{\dag}}$ and $\braket{\hat{a}^{\dag}\hat{b}}$ terms) between the two modes. Adding modes that only have classical correlations that can be unitarily removed can only add to the level of entropy.

From these observations, we analytically conclude that the entropy of the consecutive partitioning  will increase when an ``island'' that is more than 2 modes away or with more than 2 consecutive modes is added to the partitioning. We numerically show that the entropy also increases when an ``island'' that is 2 modes away with 2 or less consecutive modes is added to the partitioning in Figs.~\ref{reduction1} and~\ref{EntropyDiff1}.

\end{document}